\documentclass[intlimits,twoside,a4paper]{article}
\usepackage[cp1251]{inputenc}

\usepackage[eqsecnum]{cmpj3}

\issue{2026}{29}{1}{13401}
\doinumber{10.5488/CMP.29.13401}

\title[Surfactant solutions confined in homogeneous and Janus-like slits]%
{Surfactant solutions confined in homogeneous and Janus-like slits}
\author[T. Staszewski, M. Bor\'owko]{T. Staszewski\orcid{0000-0002-0284-4253}, M. Bor\'owko\orcid{0000-0003-1461-249X}}
\address{Department of Theoretical Chemistry, Institute of Chemical Sciences, Faculty of Chemistry, Maria Curie-Sk{\l}odowska University in Lublin, Poland}

\date{Received 4 July  2025; revised 20 October 2025; accepted 22 October 2025; published 30 March 2026}
\begin{document}

\maketitle

\begin{abstract}
We study the behavior of aqueous surfactant solutions in the bulk phase and in slit-like pores by molecular dynamics. Adsorption and self-assembly of nonionic surfactants  C$_7$E$_3$  that mimic alkyl poly(ethylene oxide)  molecules are investigated. We consider pores with the same walls and Janus-like slits. The individual walls are inert, hydrophilic, or hydrophobic. We focus on the morphology of the surfactant solution confined in different slits. The influence of a pore type and its width is discussed. The aggregative adsorption of surfactants was found. Our simulations show that in slits surfactants assemble into structures that do not occur in the bulk phases.
%
\printkeywords
%
\end{abstract}

\section{Introduction}

The use of surfactants is common in modern technology, so this problem is of great interest to researchers from both {\color{black}an applied and  fundamental perspective \cite{1,2,3,b1}.
Surfactants have been widely used in the fields of cosmetics, food, the
petroleum industry, pharmaceuticals, and  environmental remediation.  They can considerably change the surface properties of solids and hence they play a key role in many industrial processes such as dispersion/flocculation, corrosion inhibition, drug delivery, colloidal stabilization, enhanced oil recovery, and so on \cite{1,2,3,b1}.}
The fundamentals of surfactant and colloidal sciences were summarized concisely in several monographs \cite{1,2,3}.
One of the key issues is the adsorption and self-assembly of the surfactant-containing fluids on solid surfaces. Although many interesting experimental and theoretical results concerning surfactant adsorption were reported in the literature, molecular simulations play an increasingly important role in providing detailed adsorption and aggregation properties at the molecular level \cite{4}. The behavior of surfactants on different substrates was simulated, namely on flat homogeneous \cite{8,9,10,11,12} and heterogeneous \cite{13,14,15,16,17} surfaces, in pores \cite{18,19,20,21},  on nanotubes \cite{23,24,25,26,27}, and on polymer brushes \cite{28}. These studies focused on the structure of the adsorbed layer, competition between adsorption and aggregation, the excess adsorption isotherms, and many other topics.

{\color{black}Particular effort has been devoted to studying the adsorption of surfactants in pores.} The capability of a porous material to uptake and release molecules is determined by the chemical properties of the pore wall and the adsorbate, local pore curvature, and environmental conditions such as salinity, temperature, and pressure. The behavior of aqueous solutions of surfactants in porous materials is of eminent importance for technological processes involving wetting, adhesion, detergency, and mixture separation.  Understanding the principles governing the self-assembly of surfactants under confinement is the key to developing porous materials with desired properties. The behavior of surfactant-containing fluids in pores was explored primarily using computational methods.  Molecular simulations were applied to investigate the behavior of amphiphiles adsorbed in cylindrical pores \cite{18,19,20,21,22}.  It was shown that the self-assembled structure of systems involving molecules modelled as diblock polymers depends on the pore geometry and degree of confinement, that is, the relative sizes of the pore and assembling molecules  Arai et al.~\cite{18} performed dissipative particle dynamics (DPD) simulations of self-assembling short-chain surfactant inside different nanotubes. They analyzed the effect of changing the chemical characteristics of the inner wall of the nanotube on the surfactant morphologies and polymorphic transitions. The hydrophilic, hydroneutral, and hydrophobic walls were considered. Evidence was revealed for rich surfactant polymorphic structures, many of which had not been observed in bulk solutions.
Motivated by the results of small-angle neutron scattering experiments on nonionic surfactants adsorbed in nanoporous silica,  Mueter et al.~\cite{19}  determined the structure of the surfactant aggregates by DPD simulations. Their simulations reproduced the experimental findings. Furthermore, they studied the subtle interplay between aggregation and adsorption and found that increased adsorption in more hydrophilic pores leads to an increase in the effective area required by the surfactant head groups and consequently to a decrease in aggregate size. Adsorption of short-chain nonionic surfactants at the surface of mesoporous silica glass was studied by a combination of adsorption measurements and DPD simulations \cite{20,21}.  It was shown that no adsorption occurred up to the critical micelle concentration, at which a sharp increase of adsorption was observed that was attributed to ad-micelle formation at the pore walls. As the concentration was increased further, the surface excess of the surfactant decreased and eventually became negative. For wider pores excess adsorption isotherms had the azeotropic point. This was explained by the repulsive interaction between head groups, causing depletion of the surfactant in the region around the corona of the surface micelles. The effect of confinement on self-assembly and temperature-induced liquid-liquid phase separation of surfactant solutions was investigated experimentally and through all-atom molecular dynamics simulations by Wu et al \cite{22}. They studied the demixing of a model nonionic surfactant triethyleneglycol monohexyl ether in tubular nanopores of the SBA-15 silica material. They found that the nonionic surfactants showed an enhanced in situ aggregation behavior in silica pores upon increasing the temperature above the lower critical solution temperature.  

Despite intensive research, not all problems related to surfactant adsorption in pores have been fully solved. For example, the behavior of surfactants in slit-like pores has not been systematically studied.
Most computational studies concerned surfactants confined in cylindrical pores with homogeneous walls~\cite{18,19,20,21,22}. It is interesting, however, how the change of pore symmetry influences the aggregation and adsorption of surfactants. 

{\color{black}
This Special Issue is dedicated to the memory of Prof. Stefan Soko\l owski, who played an important role in advancing the study of fluids confined in slit-like pores. In the research, he used integral equation theory, density functional theory, and molecular simulations. He studied the behavior of different  fluids  in  various slitlike pores. For example, he considered adsorption of simple fluids \cite{a1,a2}, associating fluids \cite{a3}, polymers \cite{a4,a5}, electrolytes \cite{a6,a7}, amphiphilic molecules \cite{a8}, mixtures \cite{a8,a9}, and Janus particles \cite{a10} in slits. Moreover, he studied  slits with homogenous walls \cite{a1,a2},  energetically and geometrically  nonuniform pores \cite{a11}, slits with differently adsorbing walls \cite{c1}, slits filled with quenched disordered matrix \cite{a12}, pores with rough \cite{a13} or permeable \cite{a14} walls,  pores with walls modified with preadsorbed chain molecules \cite{a15},  pillared pores \cite{a16}, slits decorated with tethered polymer brushes in the form of stripes \cite{a9} and many others. Inspired by these impressive achievements, we decided to investigate the adsorption of surfactants in slit-like pores.}

In this work, we report the results of coarse grained molecular dynamics simulations of short-chain surfactants in slits. We studied a surfactant solution confined in the slits with the same walls and in Janus-like pores with different walls. The walls were inert, hydrophilic, and hydrophobic. We focused on the morphology of the investigated systems and found different aggregates adsorbed on individual surfaces or between them. The various pillars were formed. The influence of the nature of the walls and the pore width on the structure of adsorbed fluid is discussed. New structures that do not occur in bulk systems are found.

\section{Simulation methodology}
\subsection{Model}

We considered a surfactant-solvent mixture confined in a slit-like pore.  
A surfactant molecule was modelled as a diblock polymer (C$_7$E$_3$) consisting of  7 {\color{black} hydrophobic (tail) and 3 hydrophilic (head)} segments. All segments were spheres of diameter sigma. A solvent (water) molecule was a sphere of diameter $\sigma_\text{W}$.

The pore walls were crystalline lattices fcc consisting of {\color{black}beads} (S) of
diameters $\sigma_\text{S}$. The crystalline planes (001) corresponded to exposed surfaces. 
The walls were spaced apart by $H$ (pore width).

The chain connectivity was ensured by a finitely extensible nonlinear elastic (FENE)
segment-segment potential

\begin{equation} 
U_{\text{FENE}}=-\frac{k}{2}R_0^2 \ln \left[ 1- \left( \frac{r}{R_0} \right)^2 \right],
\end{equation}
where $r$ denoted the distance between the segments, $k$ was the spring constant,
and $R_0$ was the maximum possible length of the spring. The standard parameters of the
potential were assumed, $k = 30$ and $R_0 = 1.5$.

The interactions between all {\color{black}beads} (water molecules, surfactant segments, and solid molecules) were modelled via {\color{black}truncated and shifted Lennard-Jones potential proposed by Toxvaerd and Dyre} \cite{29}

\begin{equation} 
u(r)=\left\{
\begin{array}{ll}
4 \varepsilon_{ij} \left[ (\sigma_{ij}/r)^{12}-(\sigma_{ij}/r)^6\ \right] - \Delta u(r), & \ \ \ r < r_{\text{cut}(ij)}, \\
0, & \ \ \ {\rm otherwise,}
\end{array}
\right.
\end{equation}
where
\begin{equation} 
\Delta u(r)=u(r_{\text{cut}(ij)})+(r-r_{\text{cut}(ij)})\partial u(r_{\text{cut}(ij)}) / \partial r,
\end{equation}
and $r_{\text{cut}(ij)}$ was the cutoff distance, $\sigma_{ij}=0.5(\sigma_i + \sigma_j)$ 
($i,j$ = W, C, E, S), $\varepsilon_{ij}$ denoted the parameter that characterized interaction strengths between spherical species $i$ and $j$. The indices W, C, E, and S corresponded to solvent, {\color{black} the segments of surfactant tails, the segments of surfactant heads}, and the solid, respectively. The cutoff distance was used to switch on or switch off attractive interactions. For attractive interactions,   $r_{\text{cut}(ij)}=2.5\sigma_{ij}$, although for repulsive interactions $r_{\text{cut}(ij)}=\sigma_{ij}$. The gravity effect was assumed to be negligible.

We introduced the standard units defined in coarse-grained simulations. The diameter of solvent {\color{black}beads} was the distance unit,  $\sigma_\text{W}=\sigma$, the mass of a solvent {\color{black}bead} was the mass unit, $m_W = m$, the solvent-solvent energy parameter, $\varepsilon_{\text{WW}}=\varepsilon$, was the energy unit. The unit of time was  $\tau=\sigma(m/\varepsilon)^{1/2}$
Then, we used the reduced dimensionless quantities: reduced distances $l^* =1/\sigma$, reduced energies $E^* = E/\varepsilon$,  and the reduced temperature $T^*=k_{\text B}T/\varepsilon$, where $k_{\text B}$  is the Boltzmann constant.

We also defined the reduced density of the $k$-th component $\rho_k^*=\rho_k\sigma^3$  where $\rho_k=N_k/V$ denotes its number density, $N_k$  was the total number of {\color{black}beads} $k$, and $V$ is the volume of the system. The total reduced density of surfactant $\rho^*=\rho^*_\text{C}+\rho^*_\text{E}$.

We introduced the coarse-grained model since it is very flexible and can be easily adapted to different real-world systems. For example, we can assume that the solvent segment (W) corresponds
to one or two water molecules, while the nonpolar segment C contains a few CH$_2$ groups \cite{12}. {\color{black}However, a limitation of our model is that it does not explicitly take into account electrostatic interactions.
 Nevertheless, the model reproduces the basic features of the systems studied.}
The Lennard-Jones energy parameter can be modified to create a wide spectrum of interactions. In this way, it is possible to generate different surfaces, from hydrophilic to hydrophobic. {\color{black}The obtained results relate to experimental observations for nonionic surfactants on inert or hydrophobic surfaces \cite{28a}.}

In this study, we used parameters similar to those applied in previous coarse-grained {\color{black}molecular dynamic simulations of the interfacial systems involving polymers \cite{26a,26b,30,31}}.

The parameters used in the simulations are described below. To limit the number of parameters we assumed that all {\color{black}beads} had the same diameters ($\sigma_\text{E}=\sigma_\text{C}=\sigma_\text{W}=\sigma$).
Our model does not represent a specific chemical system, but rather a simple approach to easily tune the behavior of different systems and many sets of parameters.

\subsection{Simulation details}

Molecular dynamics (MD) simulations were carried out using the  LAMMPS package  \cite{32,33}. 
The Nose-Hoover thermostat was employed to control temperature, $T^* = 1$. The equations of motion were integrated using the velocity Verlet algorithm. The time step was set to $\Delta t^*=0.002$. Each system was equilibrated for at least $10^8$ time steps until its total energy reached a constant level. {\color{black}Moreover, we repeated the simulations for several replicas and obtained analogous system structures}. The production runs spanned at least $10^7$ time steps.

We investigated the surfactant-solvent mixture in the bulk system and slit-like pores. Simulations were performed in a rectangular box of reduced dimensions equal to $L^*_x= L^*_y= 40.598$ and $L^*_z=40.0$ or $L^*_z = H^*+\zeta$  along the axes $x$, $y$, and $z$, respectively. The $\zeta$ denotes the thickness of surface layers. In the case of a bulk system, standard periodic boundary conditions in all directions were assumed. However, for
the adsorption systems, the periodic conditions were applied only in the $x$ and $y$
directions. To mimic the pore walls, three layers of {\color{black}beads} S that formed the fcc crystalline lattice were located at the bottom (the first wall) and on the top (the second wall) of the box.

The OVITO 3.0.0 software was used for visualization of the equilibrium configurations \cite{34}.

\subsection{Description of studied systems}

We performed simulations of aqueous solutions of model nonionic surfactant C$_7$E$_3$ in a bulk phase and in different slit-like pores. The reduced density of the solvent was $\rho_\text{W}^*= 0.7$ \cite{28}, while the reduced surfactant density $\rho^*=0.0625$.

We assumed that self-interactions between {\color{black}beads} (WW, EE, CC) were attractive with
$\varepsilon_{\text{WW}}^*=\varepsilon_{\text{EE}}^*=\varepsilon_{\text{CC}}^*=1.0$. The cross-interactions between segments and solvent molecules were also attractive but stronger, $\varepsilon_{EW}^*=1.5$. The remaining pair-pair interactions were repulsive ($\varepsilon^*_{\text{CW}}=\varepsilon^*_{\text{EC}}=1.0$). We changed the interactions of segments E and C with both pore walls. Table~\ref{tab1} summarizes interactions with the surfaces.

We changed the interactions of segments E and C with both pore walls and the pore width.
Table \ref{tab1} summarizes the interactions with the surfaces. We studied slits with wall-separation $H^*=15$, $10$, and~$5$. We considered pores with the same and different type of walls. To make the analysis of the results easier, we introduced the following systems codes: code of the bottom wall/code of the top wall/$H^*$, for example, {\color{black}S1/S1/5} denotes the system with the same inert walls and $H^*=5$.

\begin{table}
	\caption{Characteristics of interactions with the walls, energy parameters: $\varepsilon_{\text{WS}}^*$, $\varepsilon_{\text{ES}}^*$,  $\varepsilon_{\text{CS}}^*$. The cutoff distances for repulsive and attractive interactions were $r_{\text{cut}}^*=1$, and $2.5$, respectively.}
\centering
\vspace{1mm}
\begin{tabular}{c c c c c} 
 \hline
 Code & Type & WS ($\varepsilon_{\text{WS}}^*$) & ES ($\varepsilon_{\text{ES}}^*$) & CS ($\varepsilon_{\text{CS}}^*$) \\
 \hline
 S1 & inert & rep (1.0) & rep (1.0) & rep (1.0) \\
 \hline
 S2 & hydrophilic & att (1.0) & att (2.0) & rep (1.0) \\
\hline
S3 & hydrophobic & rep (1.0) & rep (1.0) & att (1.0)\\
\hline
\end{tabular}
\label{tab1}
\end{table}

\section{Results and discussion}
\subsection{Bulk phase}

\begin{figure}
\centering
\includegraphics[width=6.0cm]{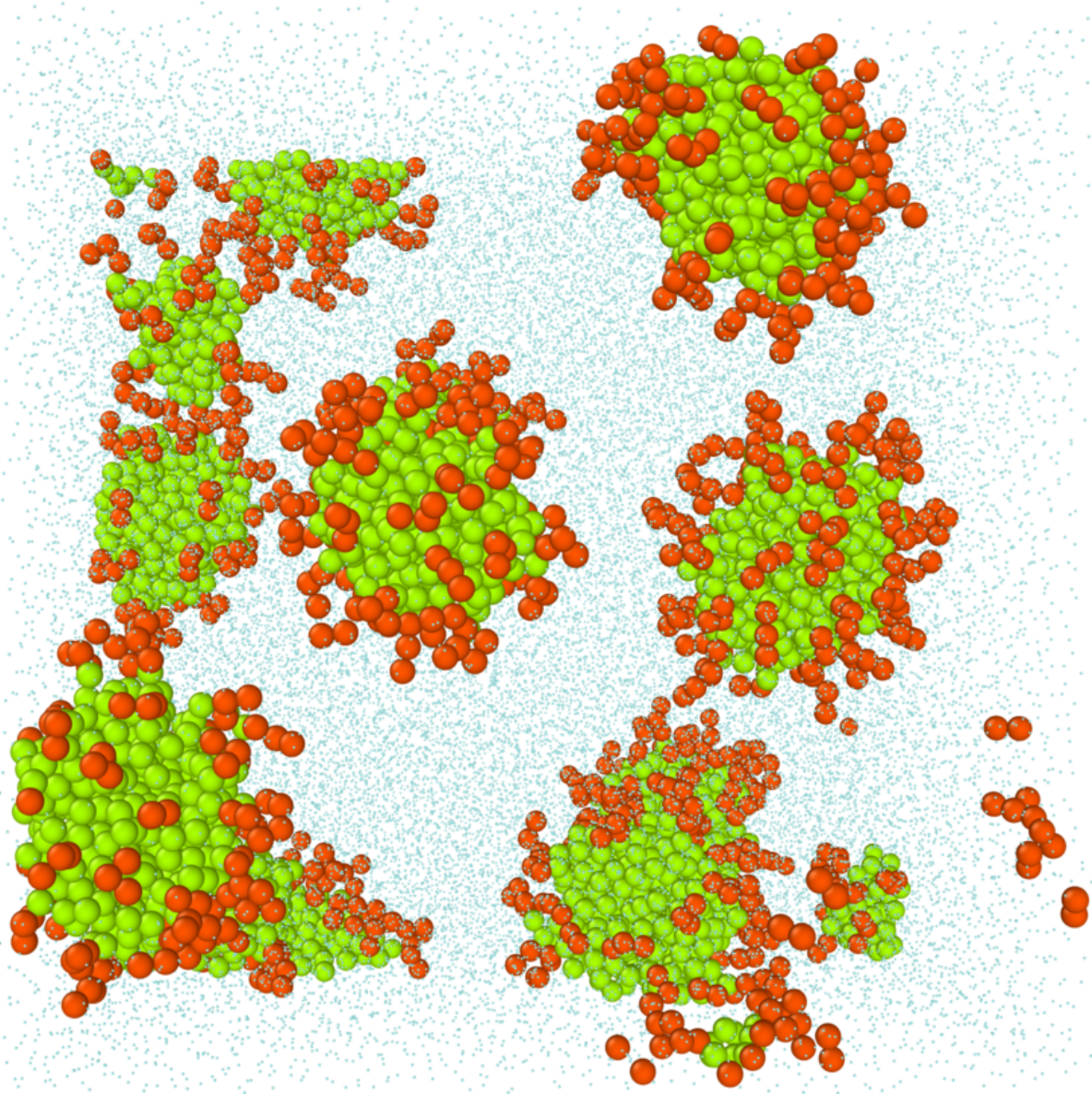}
\caption{(Colour online) Example of equilibrium configuration of aqueous C$_7$E$_3$  bulk solution. The red spheres correspond to the polar segments E, the green spheres represent the non-polar segments C, and the blue spheres are for the solvent (W). For greater clarity of the drawing, the solvent molecules are reduced.}
\label{fig1}
\end{figure}

Under the considered conditions (temperature and surfactant concentration), micelization occurred in the bulk systems. Figure \ref{fig1}  shows a typical equilibrium configuration of a bulk aqueous C$_7$E$_3$ solution.  {\color{black} Here, we see relatively large, spherical micelles and no free chains. Thus, the assumed surfactant concentration was considerably greater than the critical micelle concentration (CMC), i.e., the concentration at which micelles start to form. We have not studied the bulk systems of different surfactant concentrations.
Our goal is to} investigate the behavior of this solution inside different pores. This issue is discussed in the rest of the
paper.

\subsection{Homogeneous slits}

\begin{figure}
\centering
\includegraphics[width=6cm]{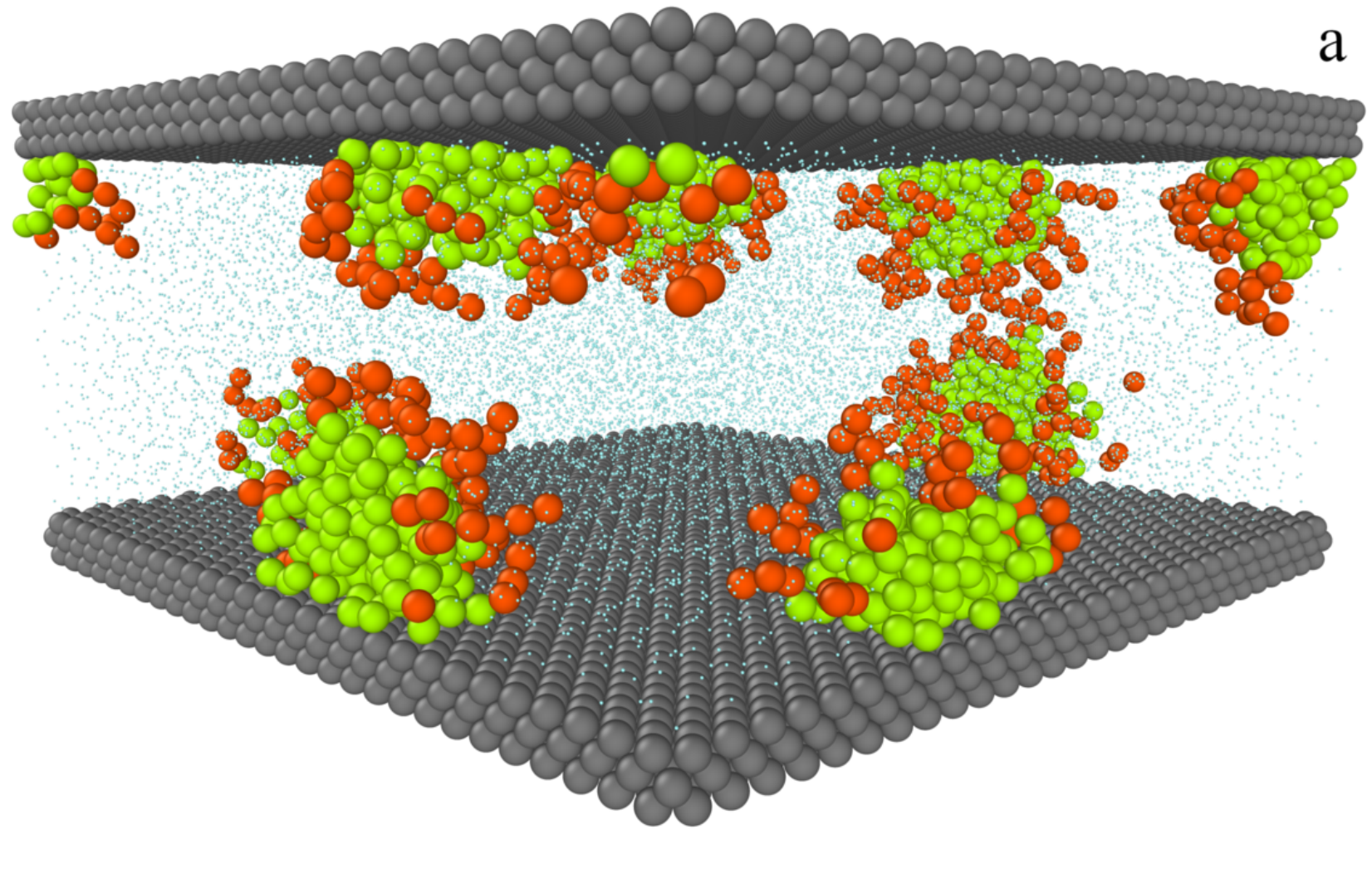}\\
\includegraphics[width=6cm]{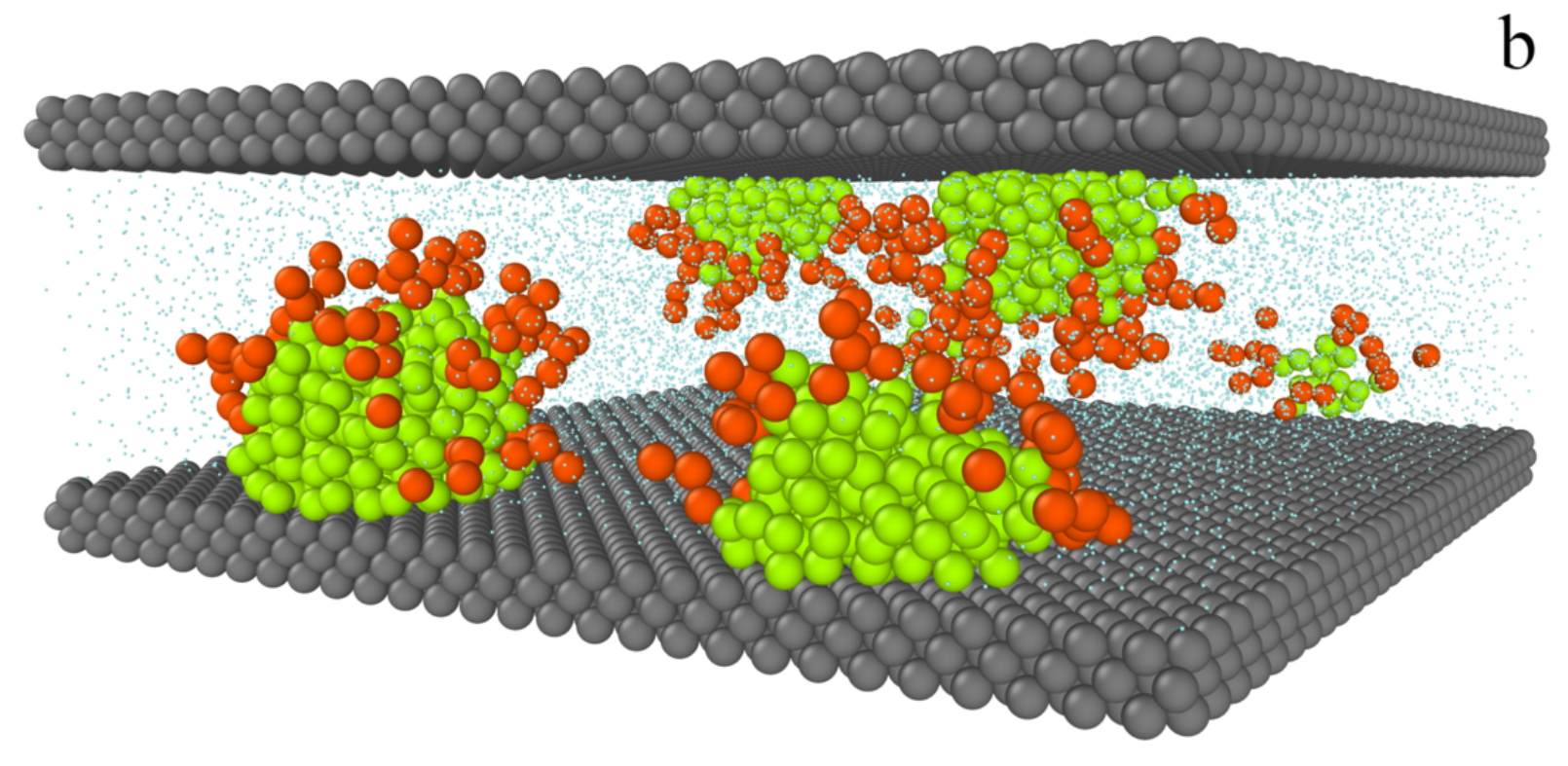}\\
\includegraphics[width=6cm]{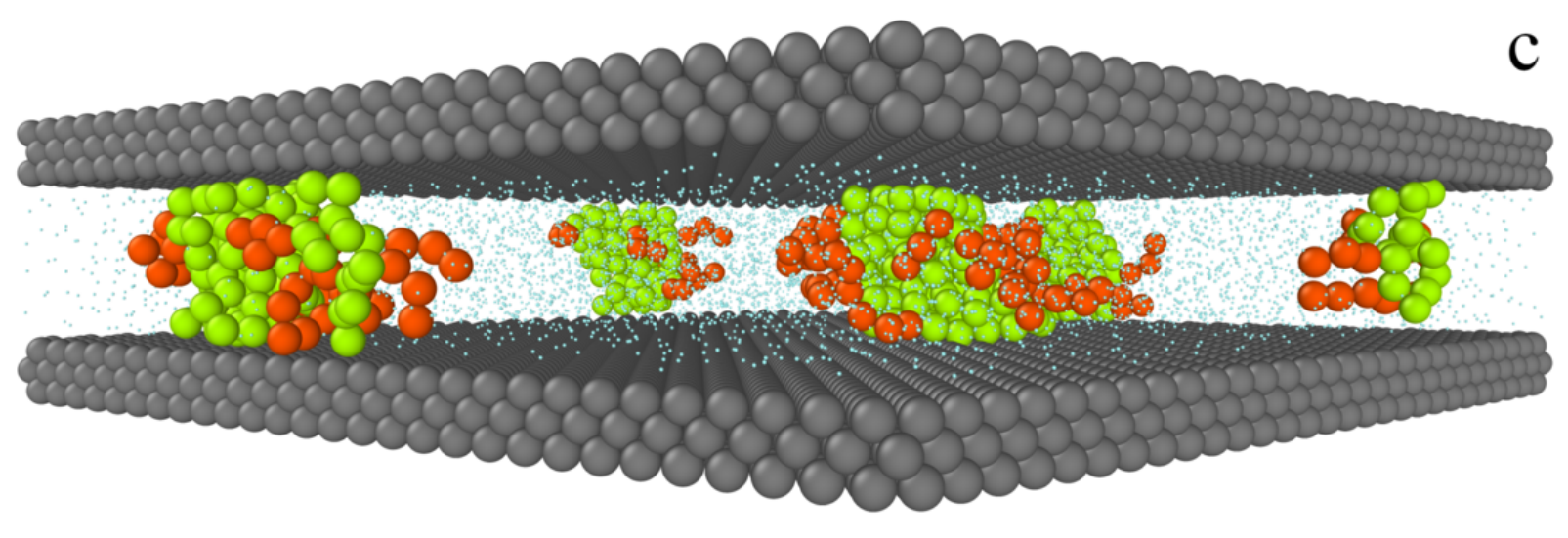}
\caption{(Colour online) Examples of equilibrium configuration of aqueous C$_7$E$_3$  solution confined in homogeneous slits with inert walls (S1) and different pore widths $H^*=$ 15 (a), 10 (b) and 5 (c).  The red spheres correspond to the polar segments E, the green spheres represent the non-polar segments C, the blue spheres are the solvent W, and the gray spheres are for the inert surface (S1). For greater clarity of the drawing, the solvent molecules were omitted.}
\label{fig3}
\end{figure}

We begin with the discussion of the results obtained for pores with inert walls (S1). Examples of the snapshots for pores with different widths are presented in figure~\ref{fig3}.  In all inert slits, surfactants strongly adsorb onto the walls.  This is the consequence of competition between interactions of surfactant molecules with the solvent and the solid surfaces.  The segments C are repelled either by the solvent molecules or by the surface. However, there are more such unprofitable contacts inside the fluid than on the walls. Therefore, the tails are accumulated on the surfaces. The adsorption is driven by hydrophobic (solvophobic) forces.  For wide slits ($H^*=15, 10$) both walls can be treated as independent adsorbents.  The adsorbed aggregates have a typical for interfaces shape of hemispheres with their tails directed towards the surface and polar groups towards the interior of the fluid (M1)\cite{1}. The same admicelles are observed in simulations carried out for hydroneutral tubes (figure~3a in \cite{18}). However,  the confinement effects are clearly visible for the narrowest pore  ($H^*=5$). In this case, the surfactants assemble into pillars that connect the walls (P1). These aggregates are considerably flattened compared to those occurring in bulk phase and in wider slits. To sum up, in the inert pores the aggregative adsorption is observed. The presence of the walls does not destroy micelles but causes their reconfiguration only.

\begin{figure}
\centering
\includegraphics[width=4.9cm]{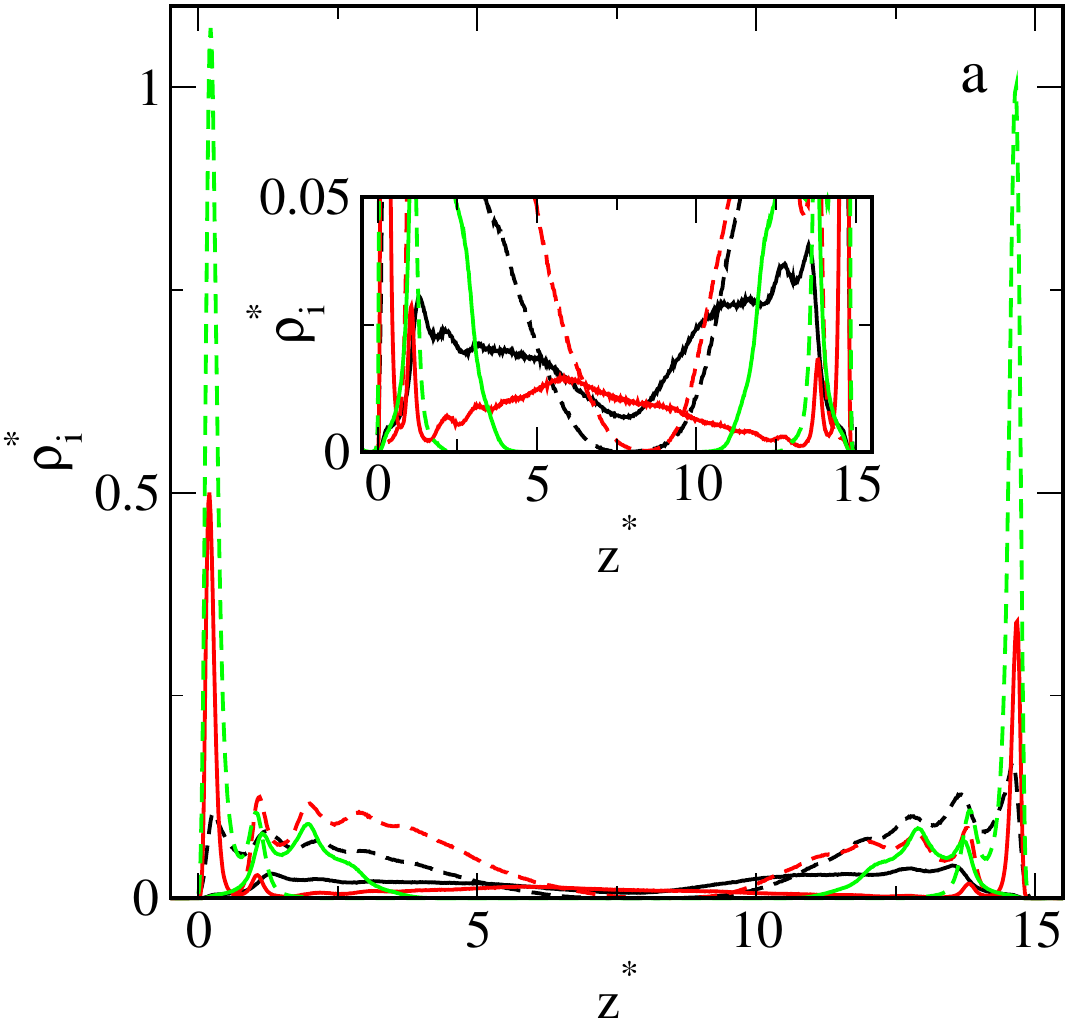}
\includegraphics[width=4.9cm]{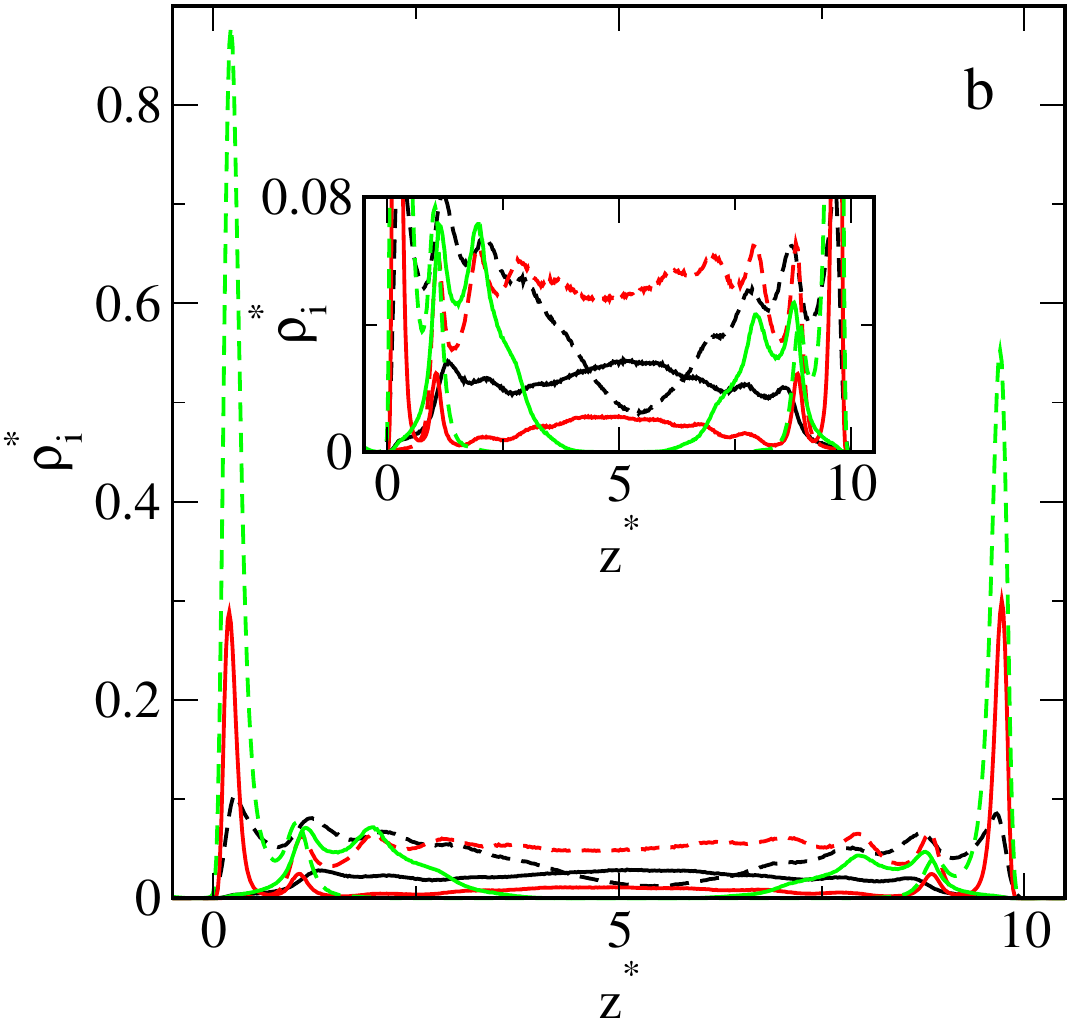}
\includegraphics[width=4.9cm]{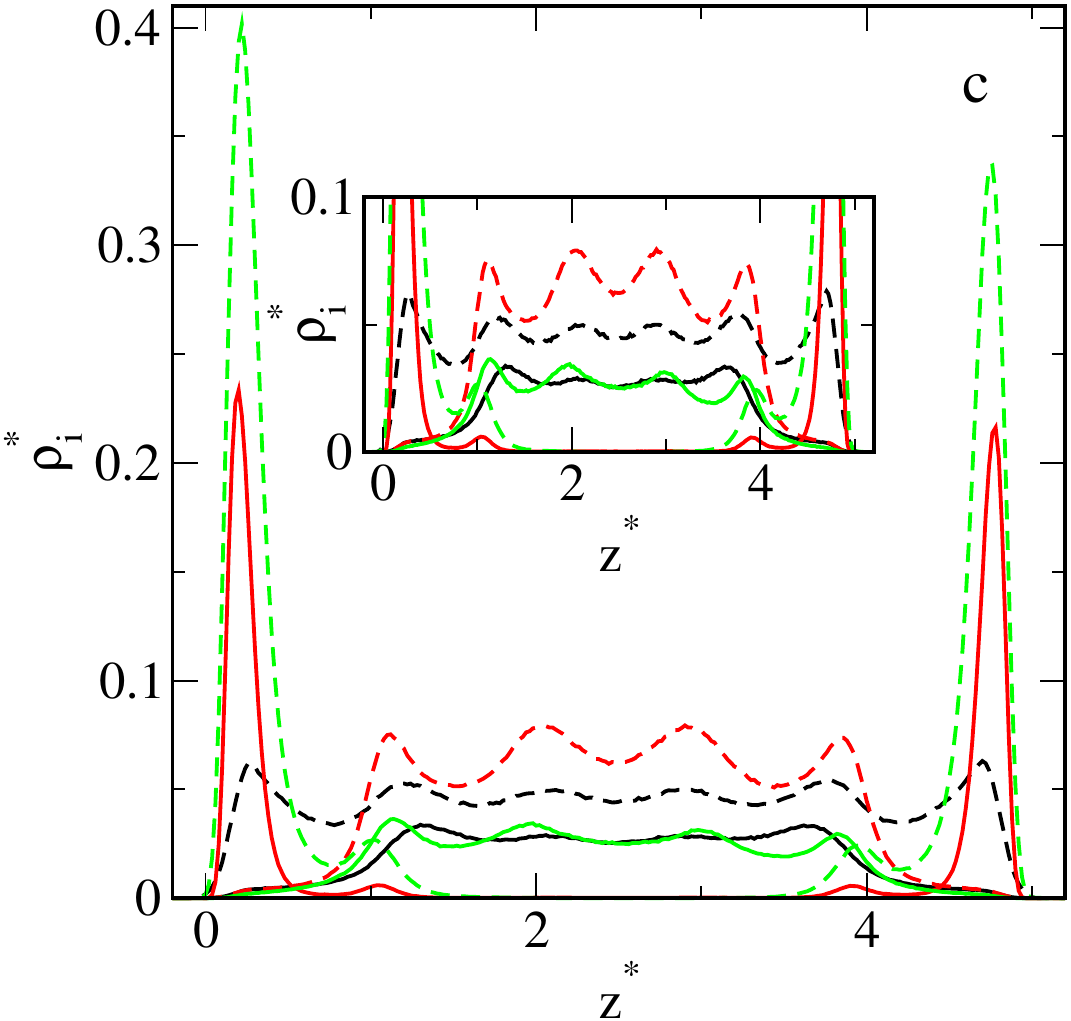}
\caption{(Colour online) The segment density profiles along the z-axis for C$_7$E$_3$  confined in homogeneous slits with different walls and different pore widths $H^*=$ 15 (a), 10 (b) and 5 (c). The densities of segments E (C) are plotted as solid lines (dashed lines).  The profiles for slits with: the inert walls (S1) are black,  the hydrophilic walls (S2) are red and the hydrophobic walls (S3) are green. {\color{black}The insets show the fragments of the density profiles with greater accuracy.}}
\label{fig2}
\end{figure}

Our findings are quantitatively confirmed by the analysis of density profiles of both kinds of segments calculated for homogeneous pores (figure \ref{fig2}). The profiles of polar (solid lines) and non-polar (dashed lines) segments are plotted for different wall separations $H^*$=15 (a), 10 (b), and 5(c). The profiles for inert walls are black. In the case of $H^*=15$, the density of segments C has three low peaks near each wall and it gradually decreases to zero as the distance from the surface increases. By contrast, the density of E-segments near the walls is close to zero, and the depletion regions are narrow.  Far away, the density increases to a wide maximum and very slowly decreases. These plots very well reflect the structure of aggregates adsorbed on isolated inert solid surfaces. The profiles estimated for  $H^*=10$ are similar to those observed in the wider inert pore. However,  the density $\rho^*_\text{C}$  in the pore center is relatively low but it does not vanish. The density $\rho^*_\text{E}$ is almost the same in the whole pore but a low maximum is observed at the pore center. This suggests that the polar groups are almost evenly distributed along the $z$-direction. In a narrower slit, the aggregates adsorbed on different walls come closer to each other. When  $H^*=5$, the profile of C-segments has low maxima on the walls while the density  $\rho^*_\text{E}$ is close to zero in this region. However, the densities of tails and heads are almost constant in the pore interior.  The profile of tails has few low humps (maxima) corresponding to successive adsorbed layers. This suggests the formation of bridges between the walls.

\begin{figure}
\centering
\includegraphics[width=6cm]{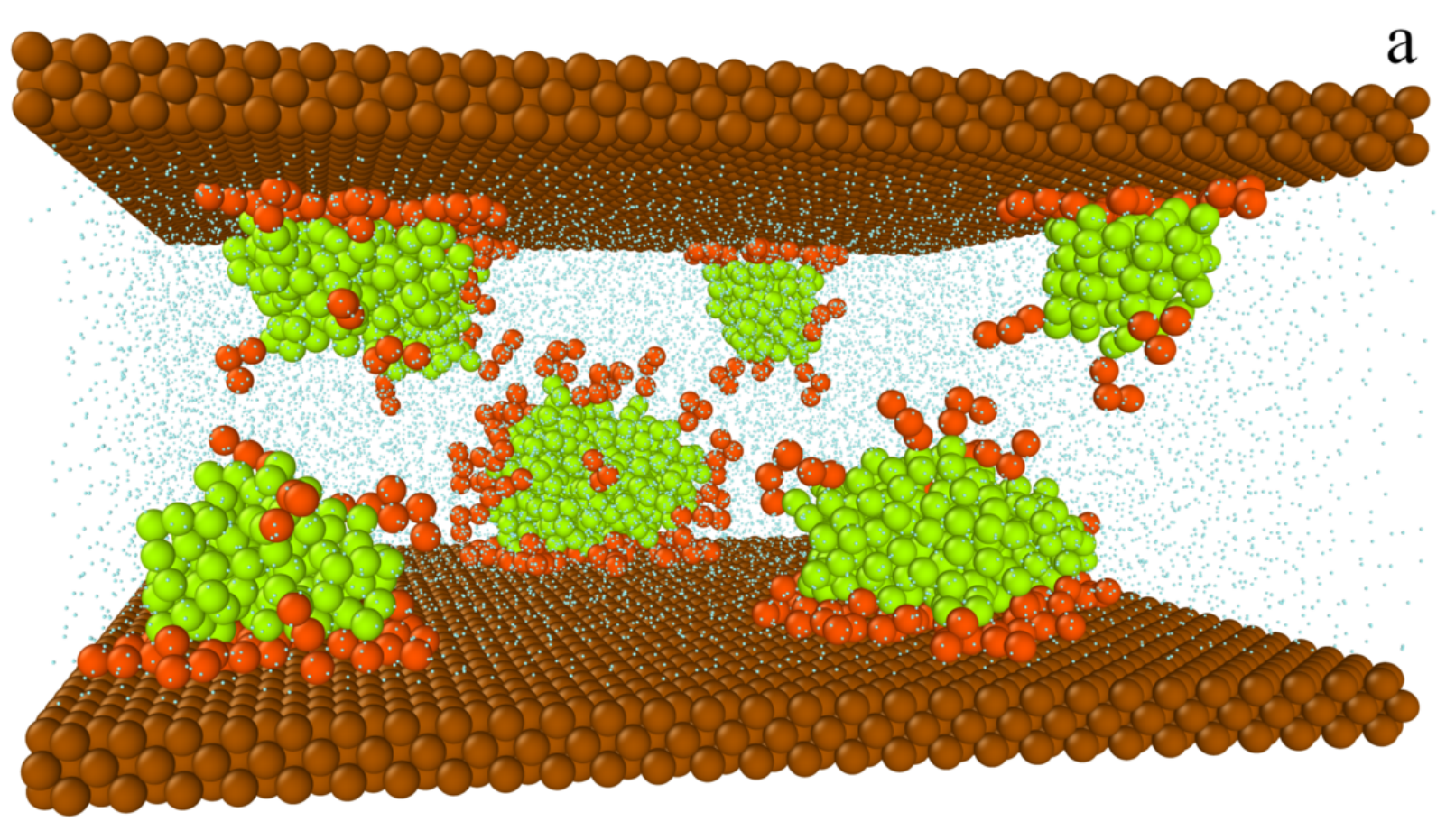}\\
\includegraphics[width=6cm]{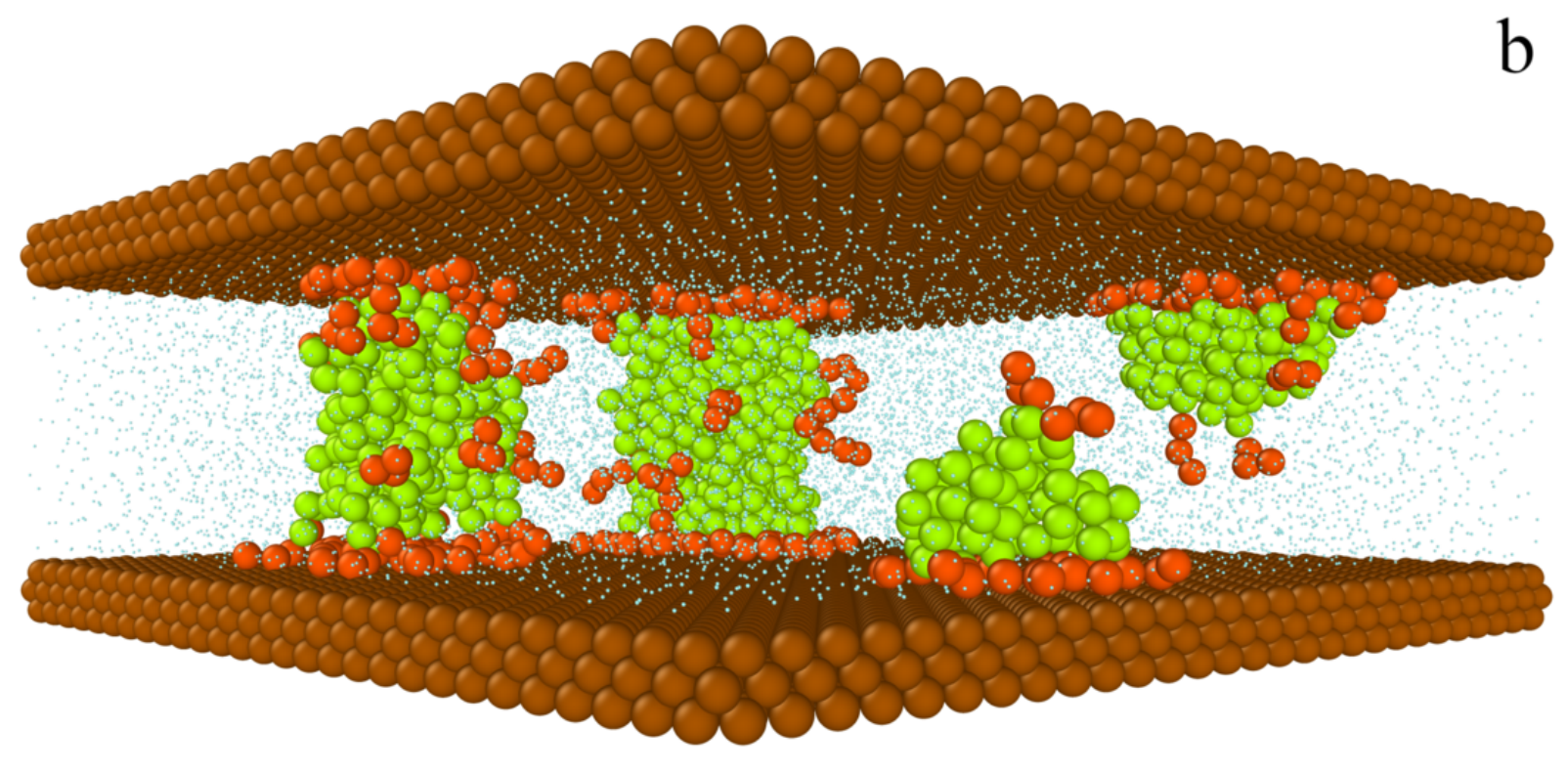}\\
\includegraphics[width=6cm]{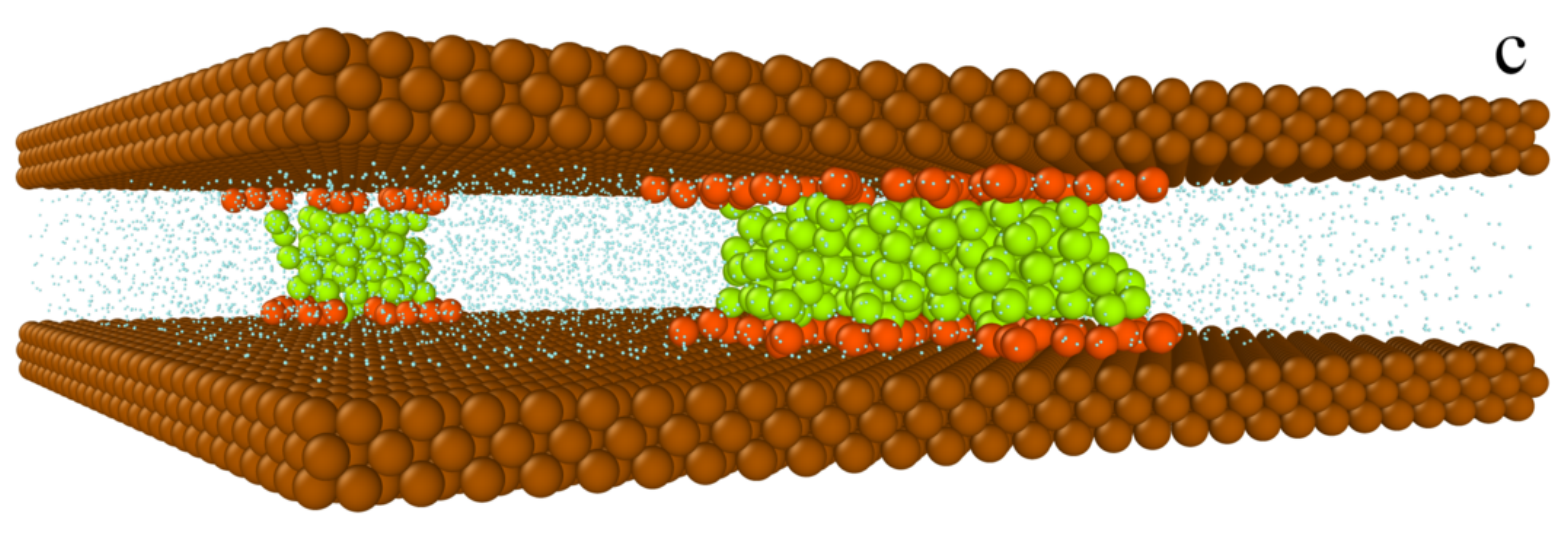}
\caption{(Colour online) Examples of equilibrium configuration of aqueous C$_7$E$_3$ solution confined in homogeneous slits with hydrophilic walls (S2) and different pore widths $H^*=$ 15 (a), 10 (b) and 5 (c). The red spheres correspond to the polar segments E, the green spheres represent the non-polar segments C, the blue spheres are the solvent W, and the brown spheres are for the inert surface (S2). For greater clarity of the drawing, the solvent molecules were omitted.}
\label{fig4}
\end{figure}

Figure \ref{fig4} illustrates the behavior of surfactants between hydrophilic surfaces (S2). In this case, the surfactants also accumulate on surfaces as micelles. However, their shape and internal structure are different. For the widest pore, the clusters resemble hats with a brim of the polar segments lying on the surface and a few E-chains on their top (M2).  When $H^*=10$, there are pillars with a pedestal of E-segments (P2) and hat-like aggregates  (M2).
The pillars are strongly elongated in the directions perpendicular to the walls. For $H^*=5$, there are only pillars observed for $H^*=5$ (P2).

The corresponding density profiles are shown in figure \ref{fig2} as red lines. We begin with a discussion of the results obtained for  $H^*=15$. The density $\rho_\text{E}^*$ has narrow and high peaks on the walls and very low peaks located somewhat far from the surfaces. In the interior of the pore, the density $\rho^*_\text{E}$  is very low but greater than zero. The distribution of the segments C is different, near each wall their density falls to zero, then three low maxima are visible and the density very slowly tends to zero. This confirms the analysis of the system structure presented above. These results are consistent with simulations performed for hydrophilic cylindrical pores and relatively low concentrations of surfactants  \cite{21}. In those systems, the segments E also accumulated near the surface while the segments C avoided it (figure 5 b, c in~\cite{21}). However, the adsorbed micelles remained spherical. Thus, the specific shape of the M2 aggregates resulted from geometric reasons related to the flatness of the surface.

In the case of $H^*=10$, the density profiles of surfactants in hydrophilic pores are similar but the $\rho^*_\text{C}$ does not decrease to zero in the middle of a slit. Indeed, there are several pillars P2. For the narrowest pore, the density profile $\rho^*_E(z^*)$  is similar to the previous. However, the density $\rho^*_\text{C}$  is high in the whole pore but on the walls. Both profiles reflect the formation of bridges between the walls.

\begin{figure}
\centering
\includegraphics[width=6cm]{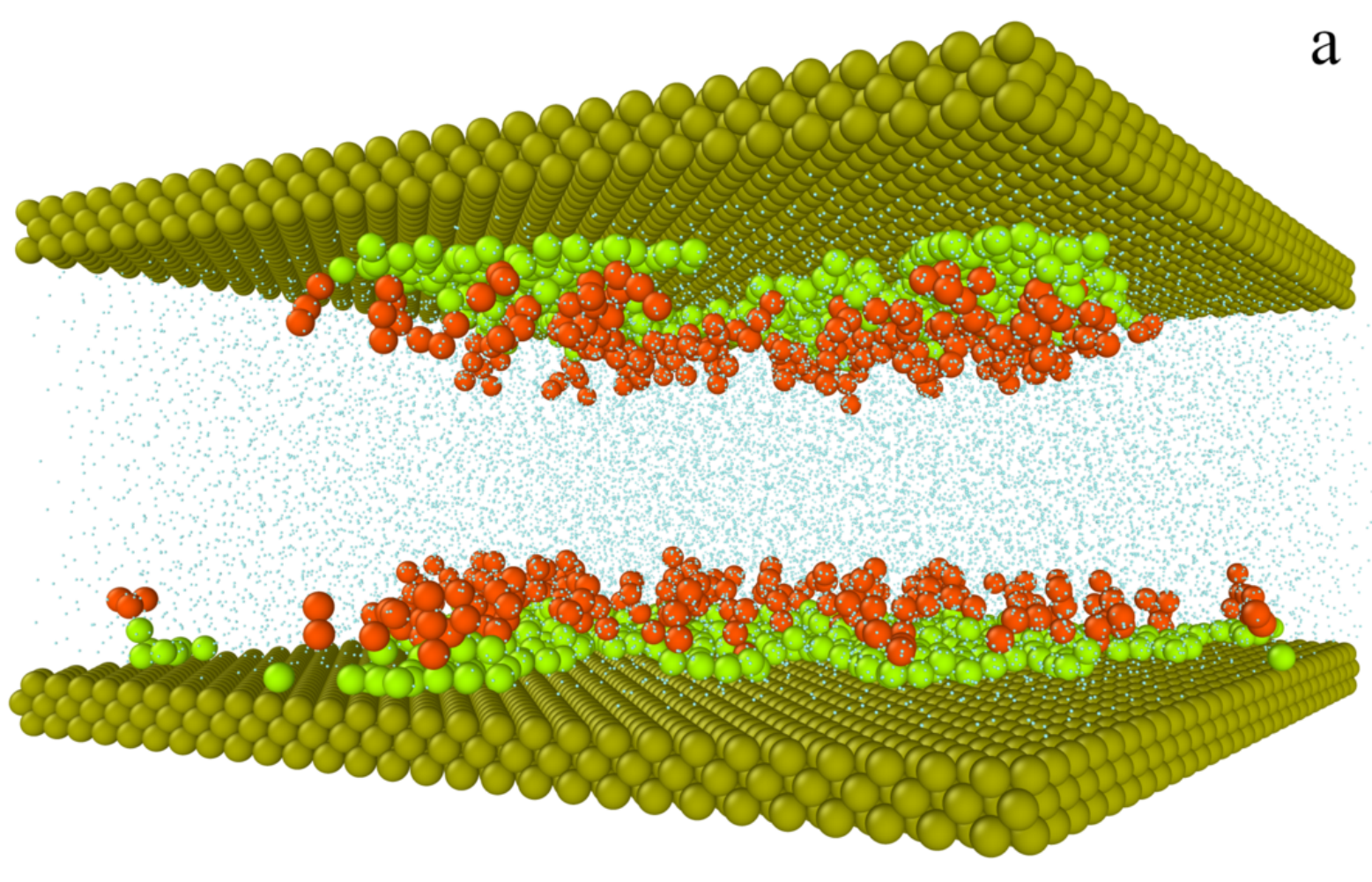}\\
\includegraphics[width=6cm]{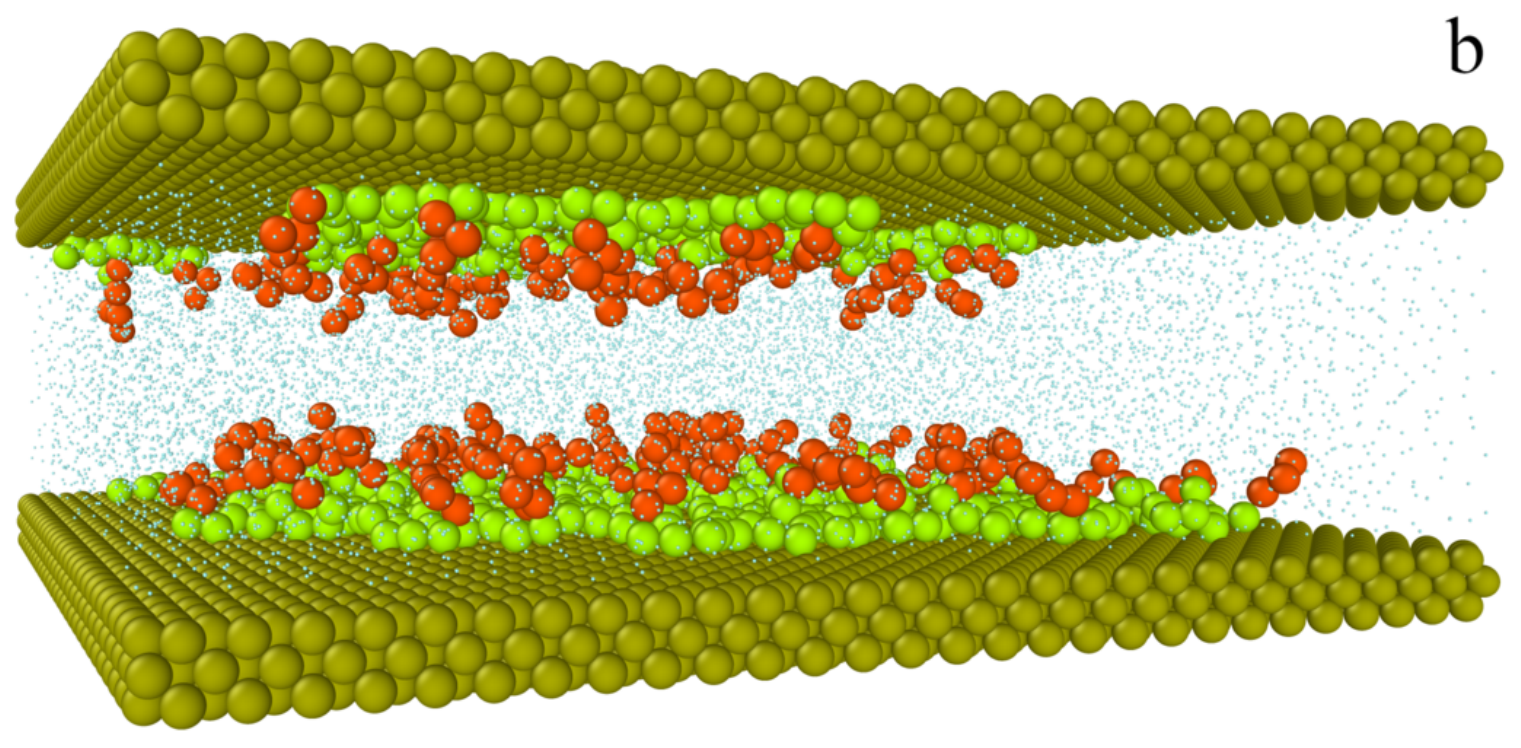}\\
\includegraphics[width=6cm]{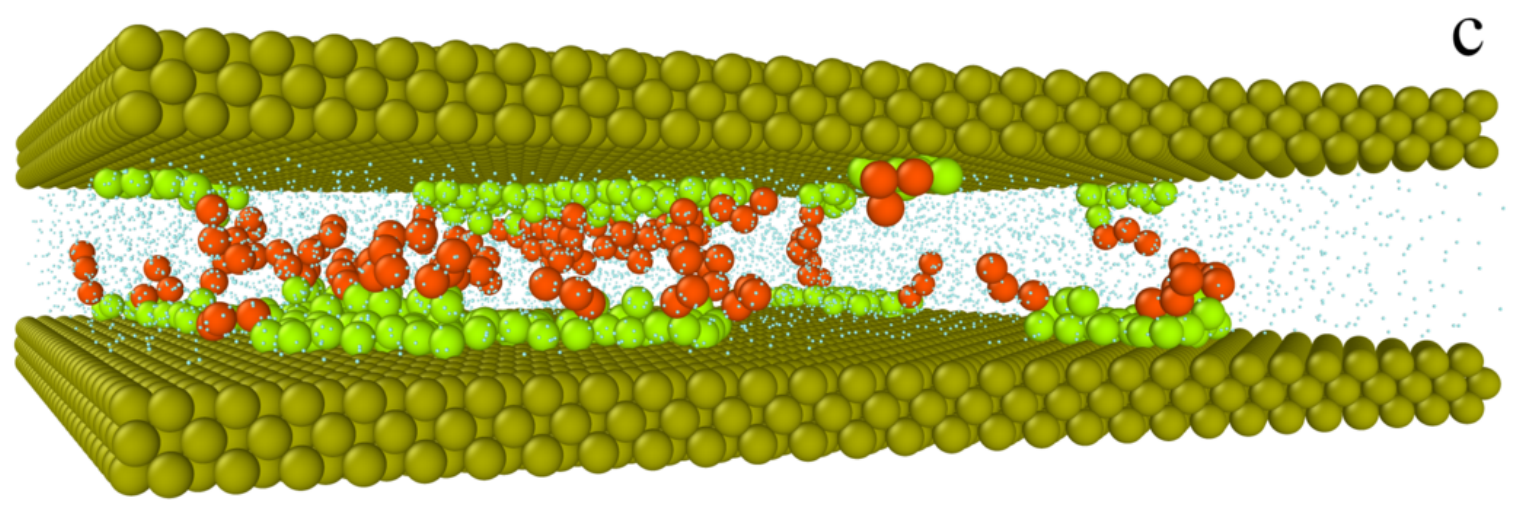}
\caption{(Colour online) Examples of equilibrium configuration of aqueous C$_7$E$_3$ solution confined in homogeneous slits with hydrophobic walls (S3) and different pore widths $H^*=$ 15 (a), 10 (b) and 5 (c).  The red spheres correspond to the polar segments E, the green spheres represent the non-polar segments C, the blue spheres are the solvent W, and the pea spheres are for the inert surface (S3). For greater clarity of the drawing, the solvent molecules were omitted.}
\label{fig5}
\end{figure}

Now, we turn to the pores with hydrophobic walls (S3). The corresponding snapshots are shown in figure \ref{fig5} while the density profiles are plotted as green lines in figure \ref{fig2}. Let us discuss the behavior of the surfactants on isolated hydrophobic surfaces ($H^*=15$). Surfactants adsorbed on surfaces as flat clusters of segments C with the E-chains directed toward the pore center (F).  The aggregates F are chaotically distributed on both surfaces. The same structure of the system is observed for narrower slits ($H^*=10, 5$). The flat aggregates do not feel the other wall. The surfactant concentration is rather low so there are too few stretching E-chains to form bridges between the walls. 

It is interesting to discuss the density profiles for these pores. The profiles of segments C have narrow and high maxima on the walls and very low maxima corresponding to the second layer of adsorbed segments. However, in the internal part of the pore, $\rho_\text{C}^*=0$. Inversely, in the immediate proximity of walls, the density of E-segments  equals zero but two wide peaks are visible near the walls. For wider pores, this density also falls to zero in the pore center. On the contrary, for $H^*=5$, the density of segments E is almost constant in the pore interior.

It should be emphasized here that an analysis of density profiles along the $z$-direction cannot clearly prove the presence of bridges between the walls. This should be confirmed by considering the segment distribution in the $x$ and $y$ directions or by monitoring snapshots.

In summary, the nature of pore walls decides the structure of admicelles on isolated surfaces. The confinement between two flat surfaces significantly influences the structure of the system only for the narrowest slits. There is one exception, namely a mixture of different aggregates appeared for hydrophilic pores of the width $H^*=10$. The observed structures are collected in table \ref{tab2}. 

The thickness of the adsorbed layers formed on the isolated surfaces is the largest for the hydrophilic wall (S2), slightly smaller for the neutral wall (S1), and the smallest for the hydrophobic surface. We also found that the number of surfactant molecules adsorbed on both surfaces was almost the same.
 
\begin{table}
	\caption{Structures observed in homogeneous slits with inert (S1), hydrophilic (S2), and hydrophobic (S3) walls and the pore width $H^*$.}
\centering
\vspace{1mm}
\begin{tabular}{c c c c c} 
 \hline
 Surface code: & S1 & S2 & S3  \\
 \hline
  $H^*=15$ & M1 & M2 & F \\
 \hline
 $H^*=10$ & M1 & M2+P2 & F \\
\hline
 $H^*=5$ & P1 & P2 & F \\
\hline
\end{tabular}
\label{tab2}
\end{table}

\subsection{Janus-like slits}

\begin{figure}
\centering
\includegraphics[width=4.9cm]{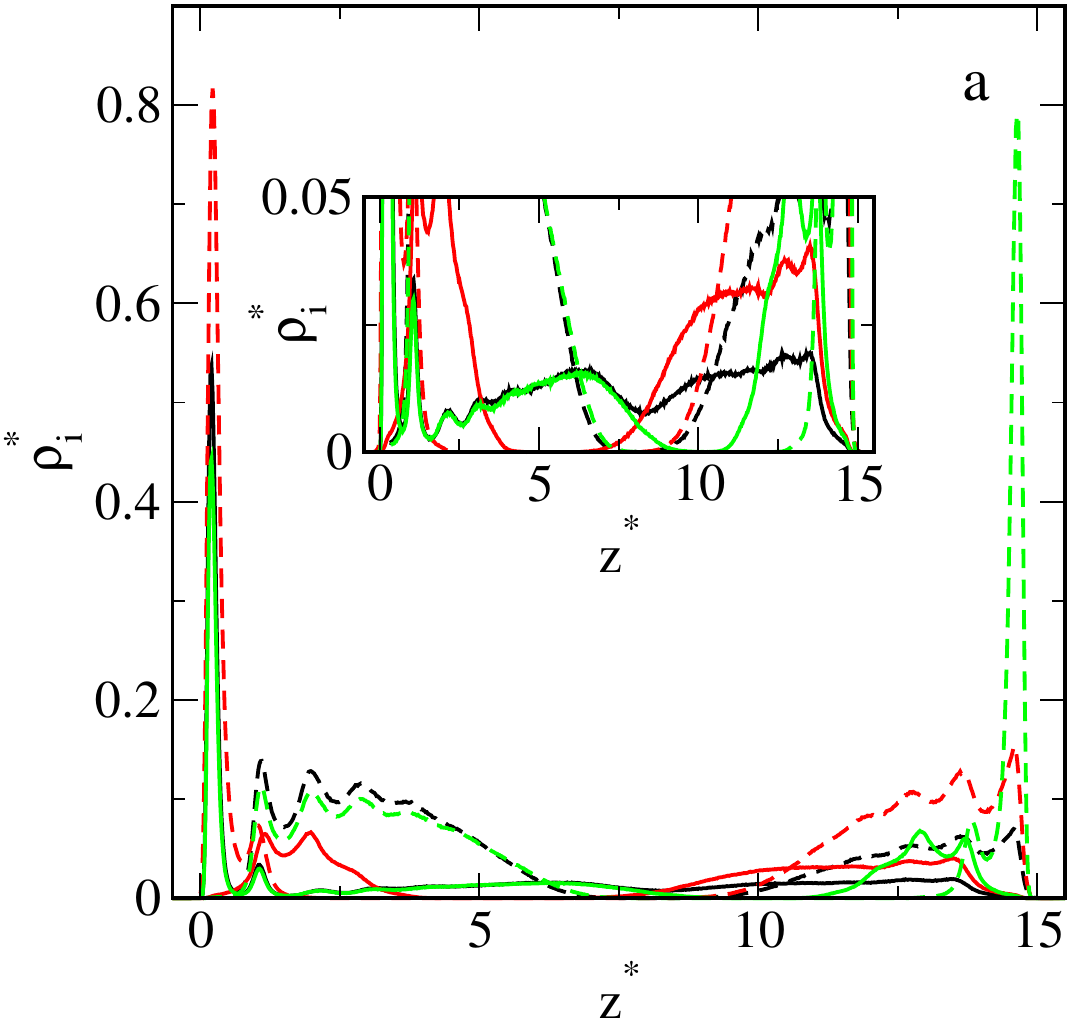}
\includegraphics[width=4.9cm]{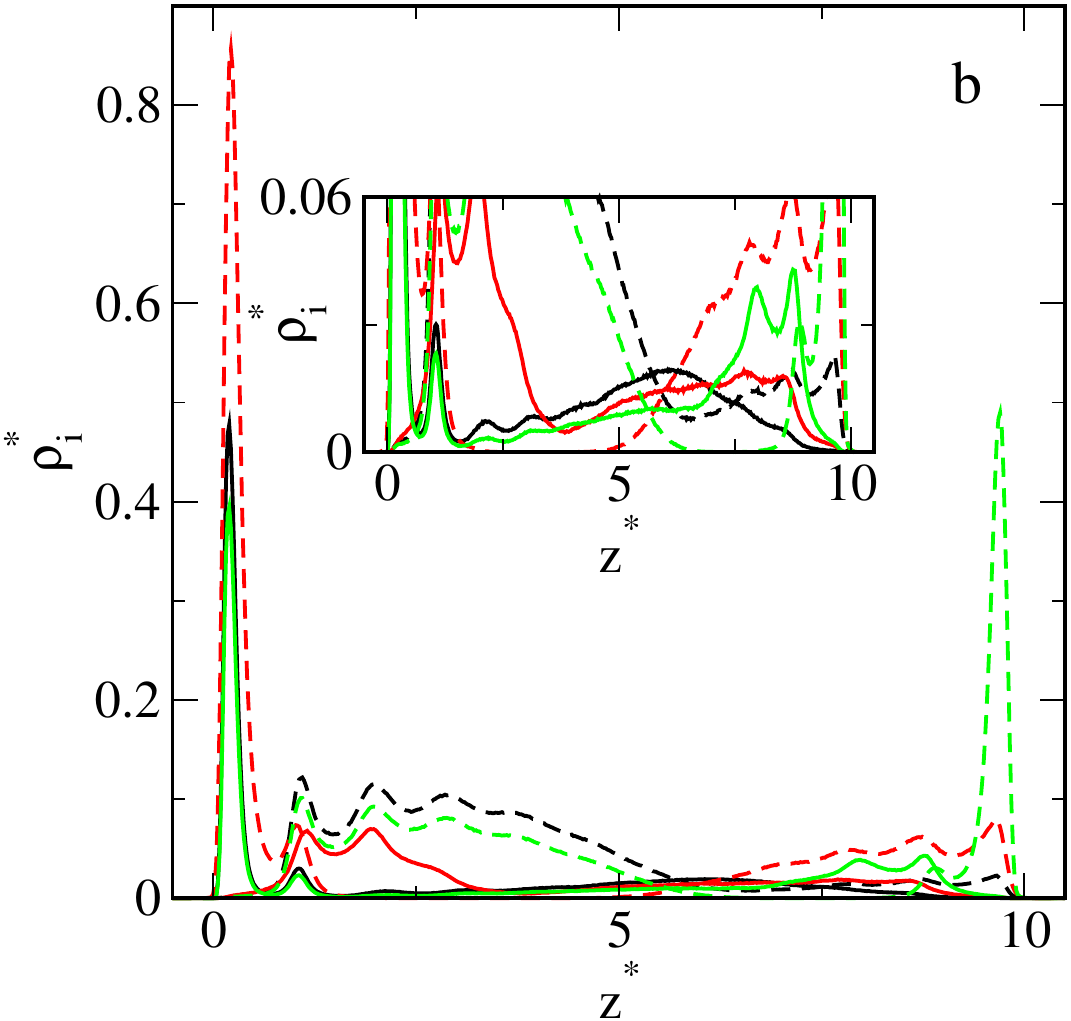}
\includegraphics[width=4.9cm]{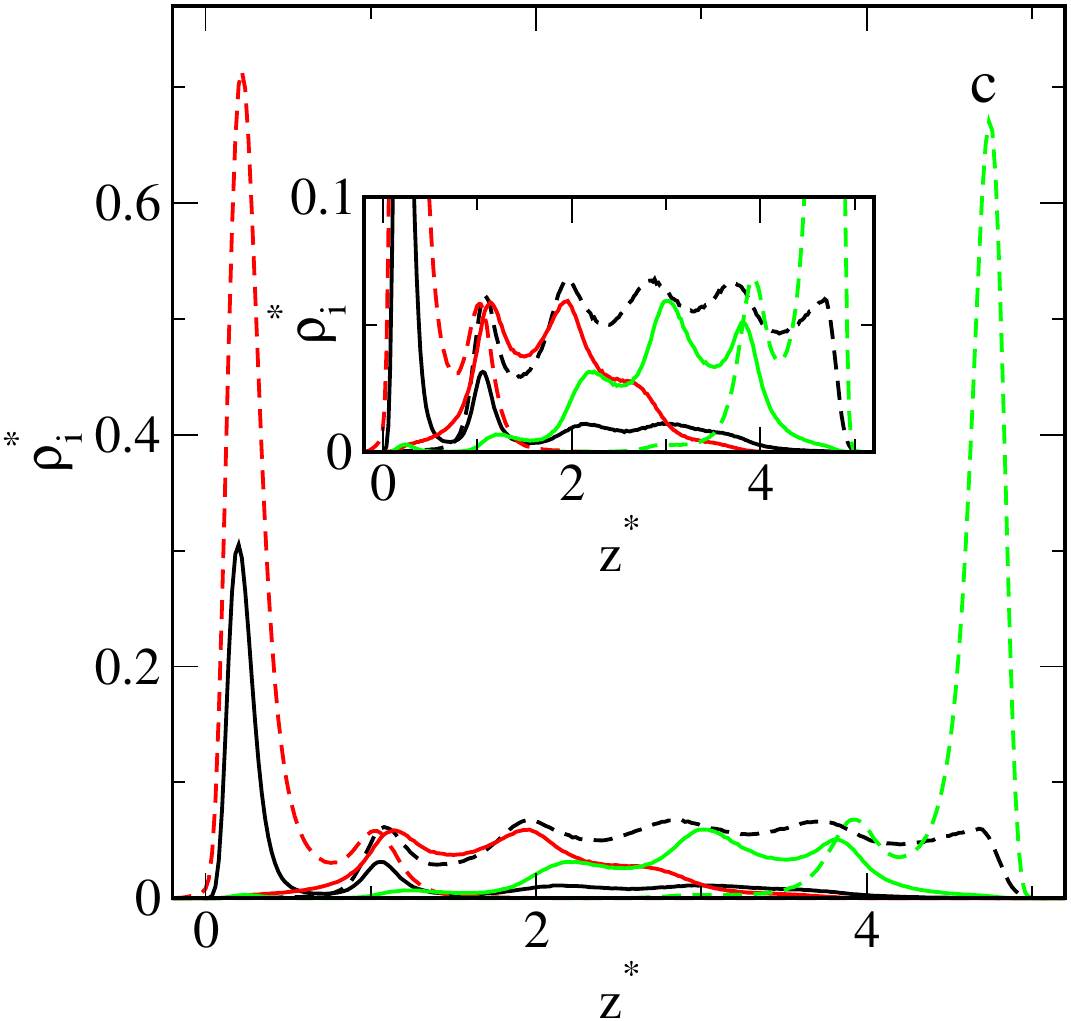}
\caption{(Colour online) The segment density profiles along the $z$-axis for C$_7$E$_3$  confined in Janus-like slits with different walls and different pore widths $H^*=$ 15 (a), 10 (b) and 5 (c). The densities of segments E (C) are plotted as solid lines (dashed lines). The profiles for slits S2/S1 are black,  for slits S1/S3 are red and for pores S2/S3 are green. {\color{black}The insets show the fragments of the density profiles with greater accuracy.}}
\label{fig6}
\end{figure}

\begin{figure}
\centering
\includegraphics[width=6cm]{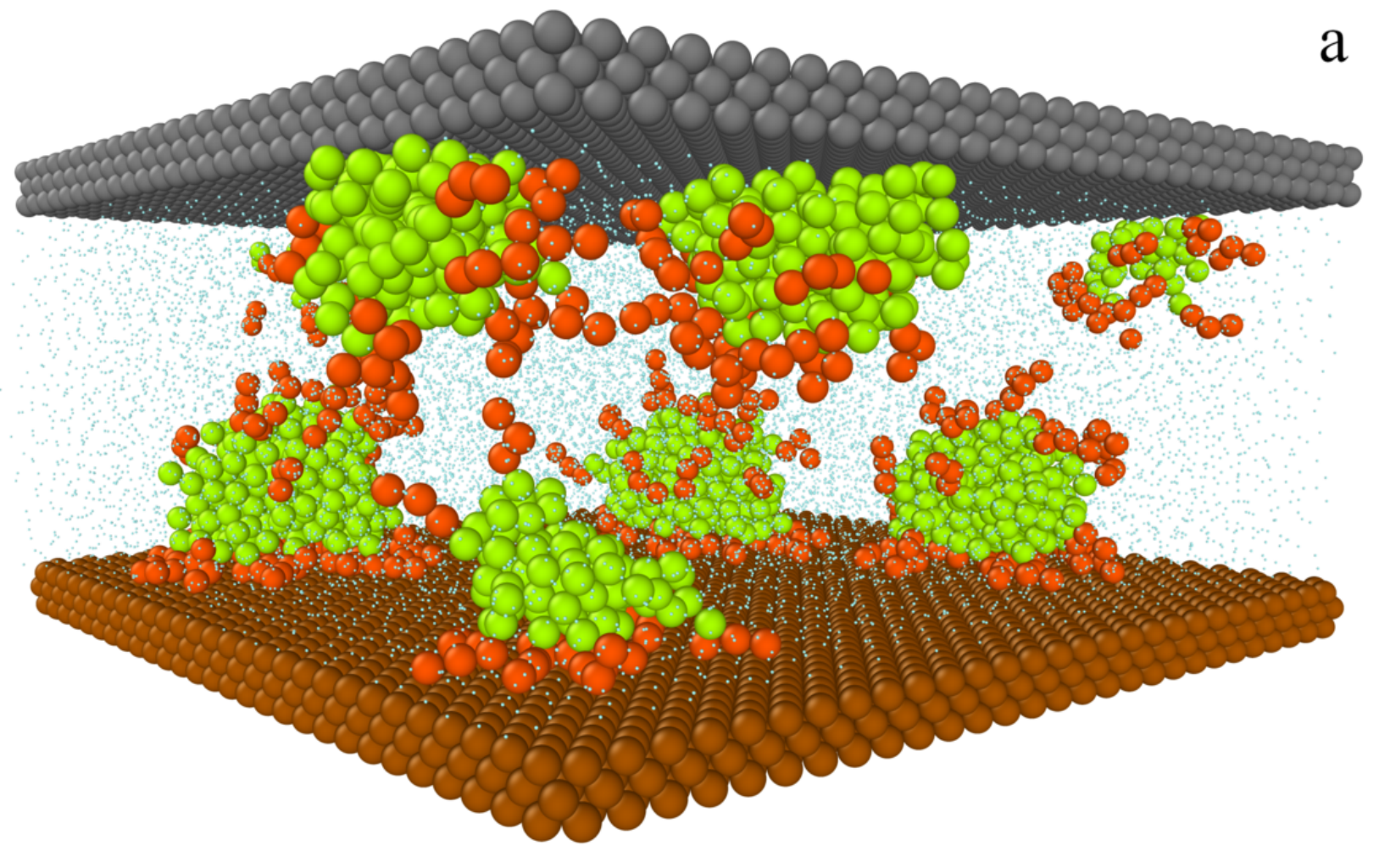}\\
\includegraphics[width=6cm]{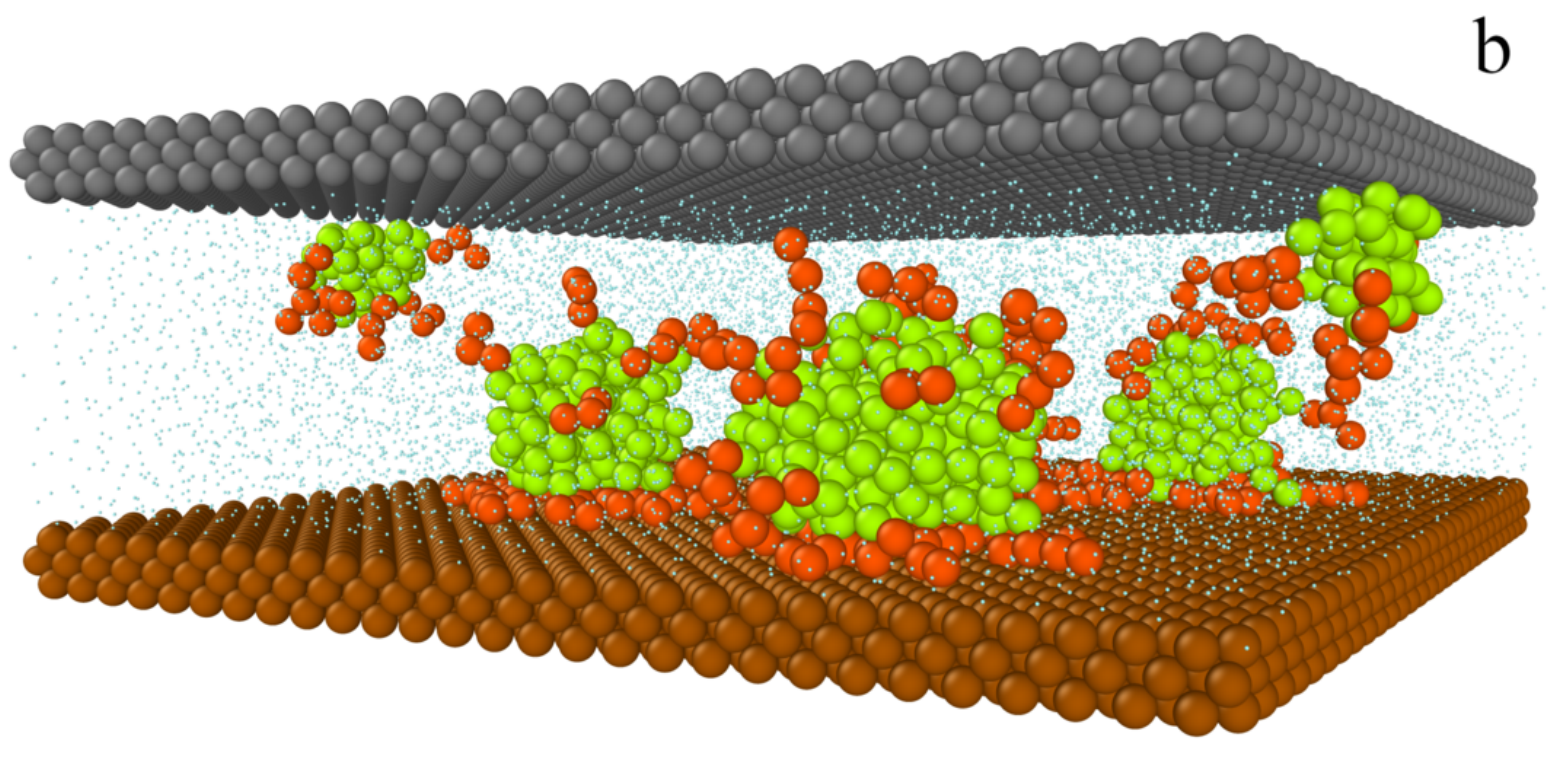}\\
\includegraphics[width=6cm]{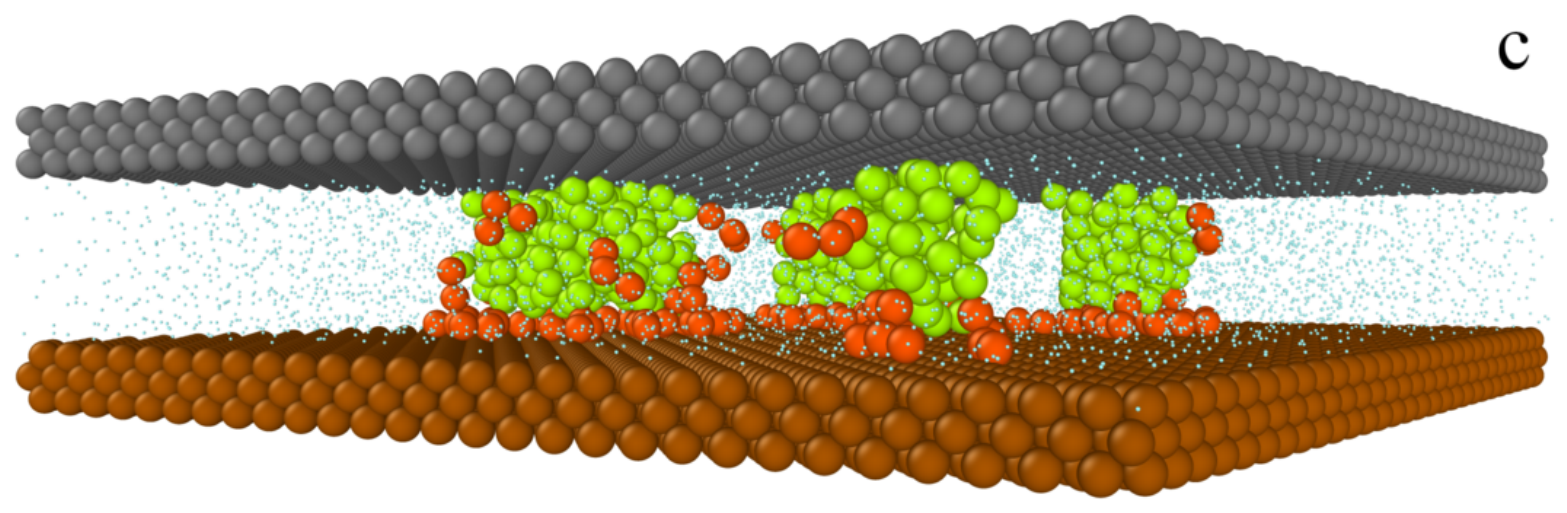}
\caption{(Colour online) Examples of equilibrium configuration of aqueous C$_7$E$_3$  solution confined in Janus-like slits S2/S1 and different pore widths, $H^*=$ 15 (a), 10 (b) and 5 (c).  The red spheres correspond to the polar segments E, the green spheres represent the non-polar segments C, the blue spheres are the solvent W, and the gray spheres are for the inert surface (S1) and the brown spheres are for the hydrophilic surface (S2). For greater clarity of the drawing, the solvent molecules were omitted.}
\label{fig7}
\end{figure}

\begin{figure}
\centering
\includegraphics[width=6cm]{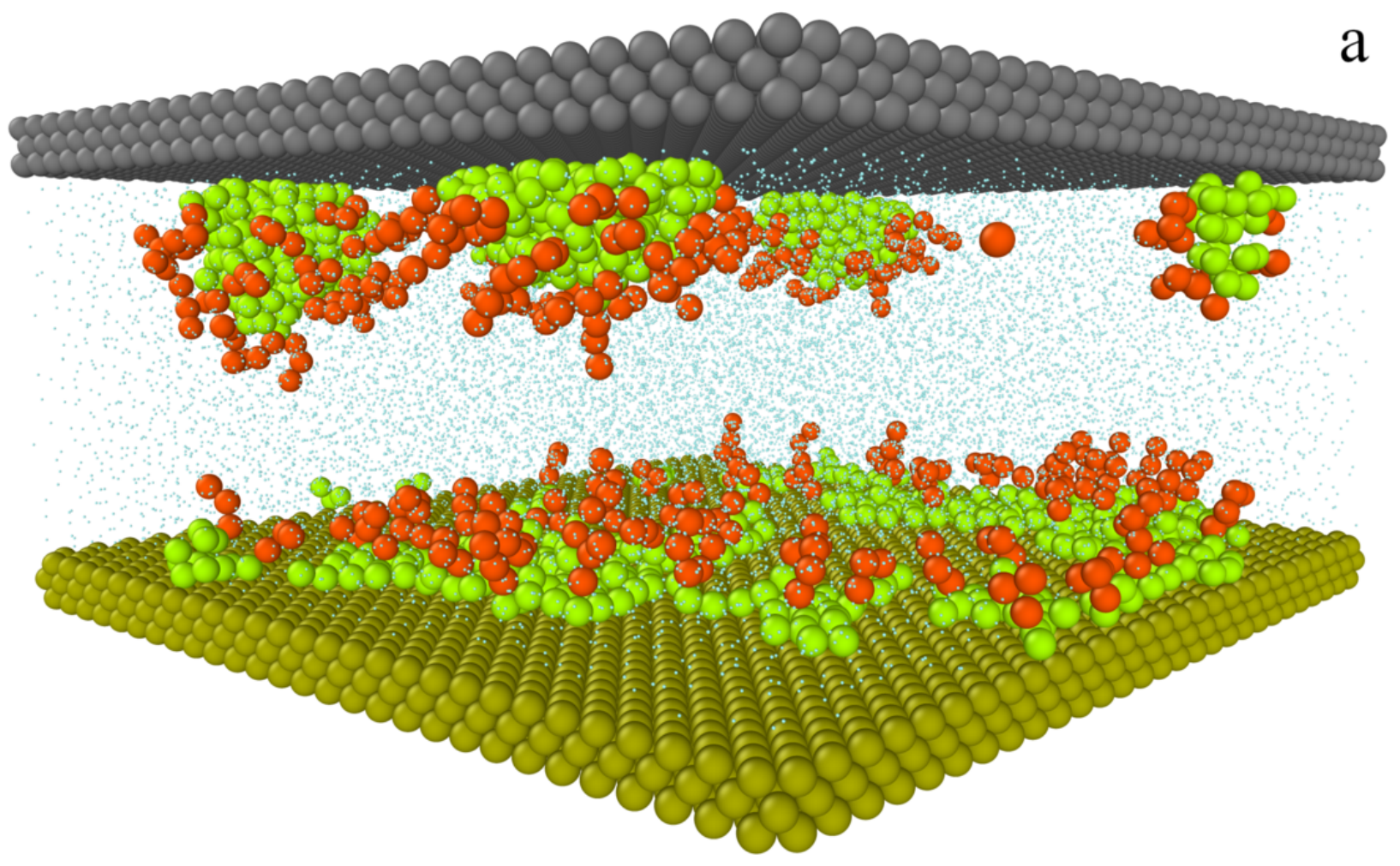}\\
\includegraphics[width=6cm]{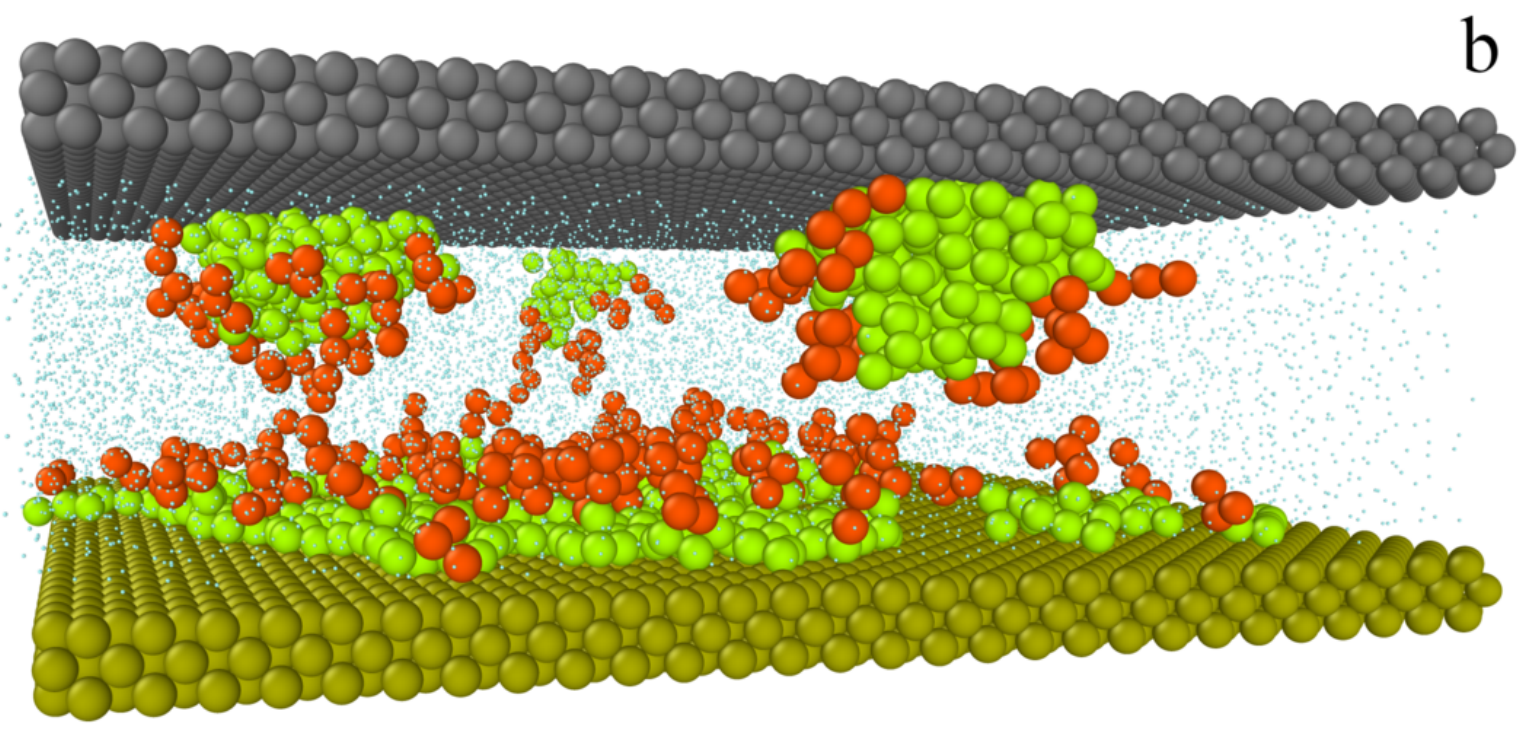}\\
\includegraphics[width=6cm]{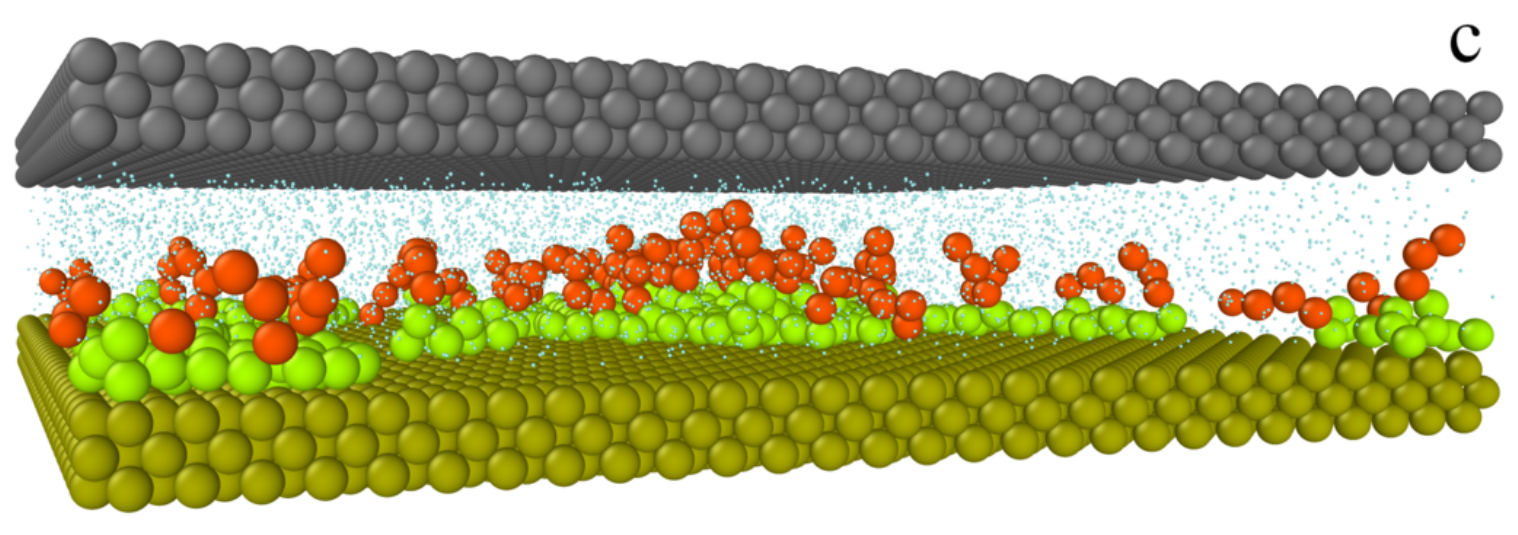}
\caption{(Colour online) Examples of equilibrium configuration of aqueous C$_7$E$_3$  solution confined in Janus-like slits S3/S1 and different pore widths, $H^*=$ 15 (a), 10 (b) and 5 (c).  The red spheres correspond to the polar segments E, the green spheres represent the non-polar segments C, the blue spheres are the solvent W, the gray spheres are for the inert surface (S1) and the pea spheres are for the hydrophobic surface (S3). For greater clarity of the drawing, the solvent molecules were omitted.}
\label{fig8}
\end{figure}

\begin{figure}
\centering
\includegraphics[width=6cm]{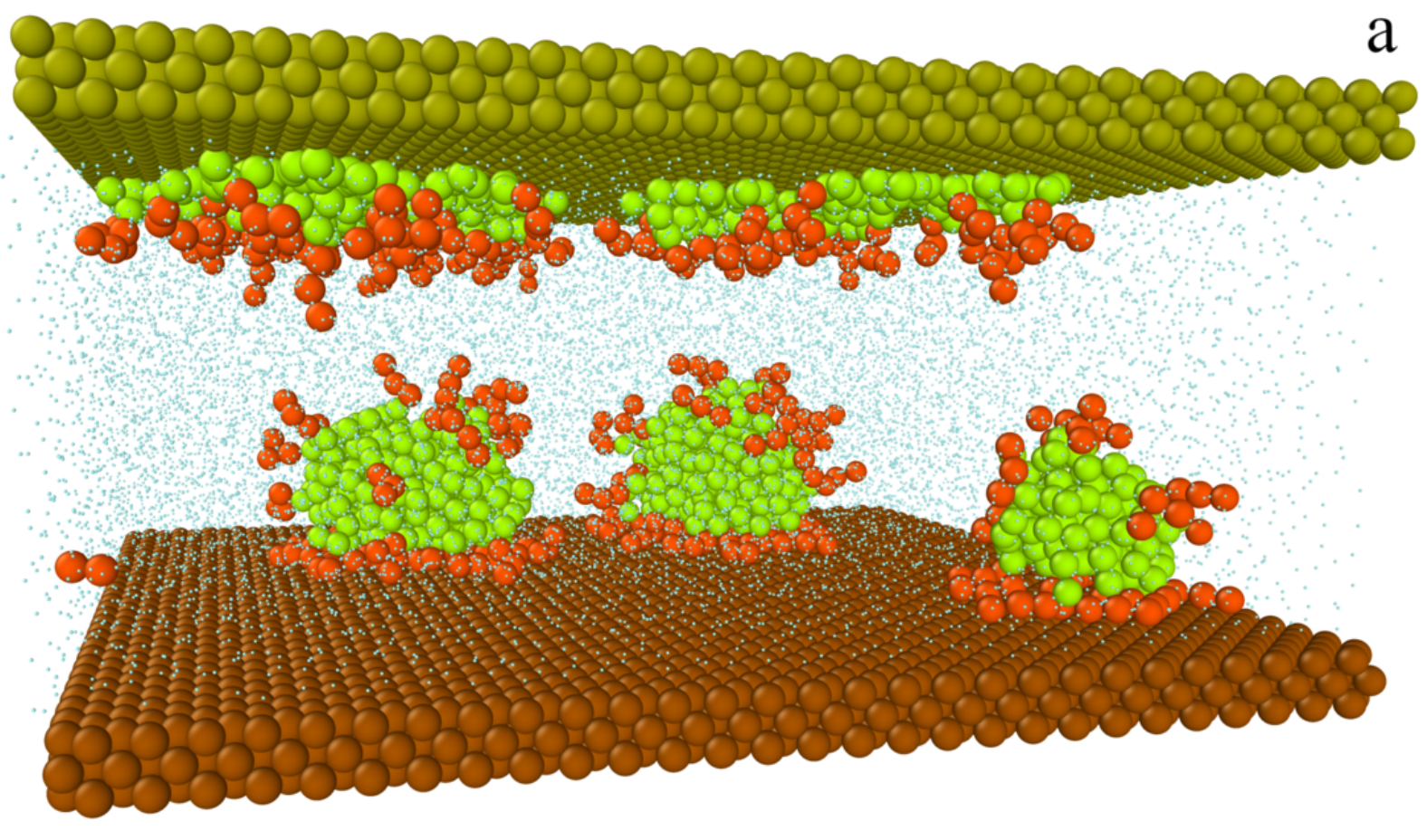}\\
\includegraphics[width=6cm]{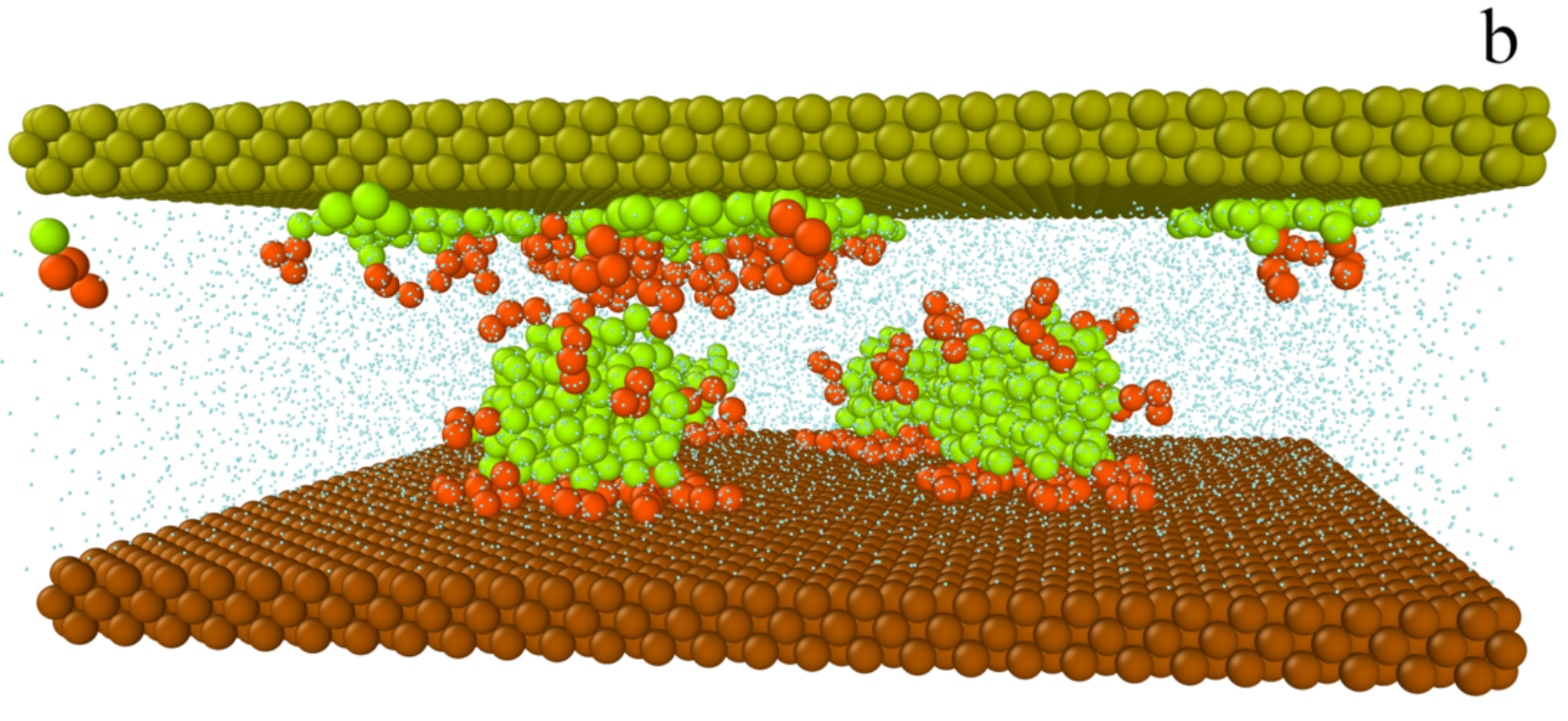}\\
\includegraphics[width=6cm]{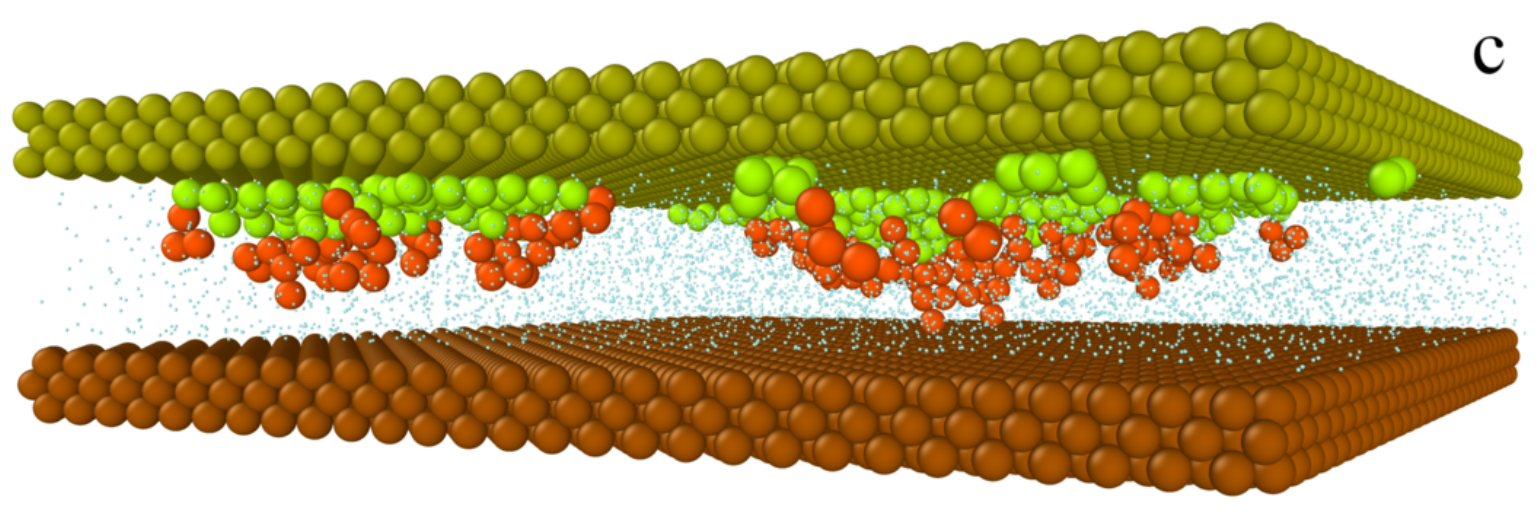}
\caption{(Colour online) Examples of equilibrium configuration of aqueous C$_7$E$_3$  solution confined in Janus-like slits S2/S3 and different pore widths, $H^*= 15$ (a), 10 (b) and 5 (c).  The red spheres correspond to the polar segments E, the green spheres represent the non-polar segments C, the blue spheres are the solvent W, and the pea spheres are for the in surface (S3) and the brown spheres are for the hydrophilic surface (S2). For greater clarity of the drawing, the solvent molecules were omitted.}
\label{fig9}
\end{figure}

We studied the behavior of surfactants C$_7$E$_3$ in three Janus-like pores with different walls: S2/S1, S3/S1, and S2/S3. The density profiles for these systems are presented in figure \ref{fig6}, while the examples of equilibrium configurations are shown in figures \ref{fig7}, \ref{fig8}, \ref{fig9}.

In wider pores, the walls behave as independent flat surfaces and surfactants adsorb in the form of aggregates occurring on the corresponding homogeneous surfaces (see parts a and b in figures \ref{fig6}, \ref{fig7}, \ref{fig8}, \ref{fig9}).

The most interesting structures were found for $H^*=5$. In the case of the pore with inert and hydrophilic walls (S2/S1), the new type of pillars was formed (P3).  The corresponding density profiles are drawn as black lines in figure \ref{fig6}c and the snapshot is shown in {\color{black}figure \ref{fig7}c.} The pillars P3 have pedestals built of E-chains adsorbed on the surface S1 and they are connected with the S2-surface via segments~C.  Moreover, a few E-chains directed toward the fluid are located near the wall S2. Such pillars were formed by the connection of aggregates M1 and M2 adsorbed on different walls. For the heterogeneous slit S3/S1, all surfactants are adsorbed on the hydrophobic surface S3 as flat clusters F (see the red lines in figure \ref{fig6}c and the snapshot in figure~\ref{fig8}c). The same behavior is observed for hydrophilic-hydrophobic pore S2/S3 (see the green lines in figure \ref{fig6}c and the snapshot in figure \ref{fig9}c).

\begin{table}
\centering
\caption{Fractions of surfactant molecules adsorbed on  individual walls $S_i$ and $S_j$ in Janus-like pore $S_i$/$S_j$.}
\begin{tabular}{c c c c c} 
 \hline
 Pore code: & S2/S1 & S3/S1 & S2/S3  \\
 \hline
 $H^*=15$ & 0.68/0.32 & 0.39/0.61 & 0.61/0.39 \\
 \hline
 $H^*=10$ & 0.87/0.13 & 0.60/0.40 & 0.64/0.36 \\
\hline
 $H^*=5$ & - & 1/0 & 0/1 \\
\hline
\end{tabular}
\label{tab3}
\end{table}

We estimated the distributions of surfactant molecules between two walls (table \ref{tab3}). In this case, the fractions of surfactant molecules adsorbed on individual walls depend on the pore type and its width.

In the pores  S2/S1, adsorption on the hydrophilic walls is greater. When $H^*$ decreases from 15 to 10, the fraction of surfactants accumulated on this surface increases. In the narrowest slit, the pillars P1 are connected with both walls. One can say that attractive interactions with a hydrophilic surface dominate in such systems.

A different relation is observed for pores S3/S1. In the widest slit, adsorption on the hydrophobic wall (S3) is greater than on the inert surface (S1). Inversely, when $H^*=10$ more surfactants adsorb on the inert surface, and for $H^*=5$, all surfactant molecules ``fall'' to this wall. We see here the competition between two forces acting on surfactant molecules, the attraction by the wall S3 and the hydrophobic repulsion toward the inert surface S1.

In the wider pores S2/S3, the surfactants adsorb mainly on the hydrophilic wall, and the distribution of these molecules between the walls is almost the same for $H^*=15$ and $H^*=10$. In the narrower slit, a change occurs, all the surfactant accumulates on the hydrophobic surface of S3.  In this case, geometric limitations play a key role, the aggregates M2 simply cannot be formed so all molecules stick to the surface S3.

Simulations have shown that the adsorption and self-assembly of surfactants in Janus-like pores are very complex processes that are influenced by the relationship between numerous parameters of the system. 

\section{Conclusions}

We employed molecular dynamics to study the behavior of aqueous surfactant solutions of a selected concentration in the bulk system and in various slit-like pores. We considered the nonionic surfactant C$_7$E$_7$ which mimicked alkyl poly(ethylene oxide). Each molecule was modelled as a diblock polymer consisting of nonpolar segments C (tail) and polar segments E (head).  The surfaces were represented by frozen S beads. We studied the pores with the same and different walls. The investigated homogeneous pores had inert (S1), hydrophilic (S2), and hydrophobic (S3) walls. We also consider three Janus-like pores: S2/S1, S3/S1 and S2/S3. Furthermore, we changed the pore width from this corresponding to isolated surfaces ($H^*=15$), through average wall separation ($H^*=10$) to the narrowest pore ($H^*=5$).

Our simulations showed the formation of relatively small spherical micelles in the bulk system under assumed thermodynamic conditions (system density, surfactant concentration, and temperature). 

The model surfactant molecules are adsorbed on all isolated surfaces in the form of different aggregates. In the case of inert surfaces, the accumulation of surfactant was caused by the hydrophobic pushing out the tails from the solvent. Adsorption on hydrophilic and hydrophobic surfaces occurs as a result of the attraction of their ``heads'' or ``tails'', respectively, by these walls. The shape and structure of adsorbed aggregates depended on the type of walls. 
The confinement also had an impact on the morphology of the surfactant solution inside pores. In sufficiently narrow pores, different pillars between the walls were found. The observed structures are collected in table~\ref{tab2}.

We also studied the adsorption of surfactants and their self-assembly in Janus-like pores. In wide pores,  surfactants were adsorbed in the form of ad-micelles occurring on the corresponding homogeneous surfaces. However, for the narrowest slit with hydrophilic and inert walls (S2/S1), new type of pillars was formed. It is interesting that for $H^*=5$ and the pores S3/S1 and S2/S3, all surfactants adsorbed on hydrophobic walls.

The behavior of surfactants was quantified by the density profiles of different segments and supplemented by the analysis of the snapshots. We also estimated the fractions of surfactant molecules adsorbed on individual walls (table~\ref{tab3}).

Our simulations were performed for one value of surfactant concentration.  Further research could concern the influence of surfactant concentration on the structure of the fluid adsorbed in slits.

Surfactant solutions confined in nanopores are an issue of great importance.  The morphology of such systems is the result of adsorption interactions, symmetry breaking, and confinement-induced loss of entropy. The complex interplay between adsorption and self-assembly plays a fundamental role in the processes occurring in confined surfactant solutions. It is possible for surfactants to self-assemble into structures that are not present in the bulk.

\section*{Acknowledgements}

We dedicate this work to the memory of Professor Stefan Soko\l owski, our mentor, collaborator, and friend.

\bibliographystyle{cmpj}
\bibliography{bibliography}

\ukrainianpart

\title{Розчини поверхнево-активних речовин, утримувані в однорідних та янус-подібних щілинах}
\author{T. Сташевскі, M. Борувко}
\address{Кафедра теоретичної хімії, Інститут хімічних наук, Хімічний факультет, Університет Марії Склодовської-Кюрі, 20031 Люблін, Польща}

\makeukrtitle

\begin{abstract}
	\tolerance=3000%
	Ми вивчаємо поведінку водних розчинів поверхнево-активних речовин (ПАР) в об'ємній фазі та в щілиноподібних порах методом молекулярної динаміки. Досліджується адсорбція та самозбірка неіоногенних ПАР C$_7$E$_3$, що імітують молекули алкілполіетиленоксиду. Розглядаються пори з однаковими стінками та щілинами типу Януса. Окремі стінки є інертними, гідрофільними або гідрофобними. Ми зосереджуємося на морфології розчину ПАР в різних щілинах. Обговорюється вплив типу пори та її ширини. Було виявлено агрегаційну адсорбцію ПАР. Проведене моделювання показує, що в щілинах ПАР збираються у структури, які не зустрічаються в об'ємних фазах.

	\keywords неоднорідні рідини, поверхнево-активні речовини, молекулярна динаміка
\end{abstract}

 \end{document}